\documentclass[paper=a4,fleqn,usenatbib]{mnras}
\usepackage{wrapfig}
\usepackage{footnote}
\usepackage{newtxtext,newtxmath}
\usepackage{amsmath}
\usepackage{graphicx}
\usepackage{epstopdf}
\usepackage{multirow}
\usepackage{natbib}
\usepackage{afterpage}
\usepackage[T1]{fontenc}
\usepackage{ae,aecompl}
\usepackage{upquote}
\usepackage{float}
\title[Go with the Flow]{Go with the Flow: \\ Understanding Inflow Mechanisms in Galaxy Collisions}

\author[K. A. Blumenthal et al.]{
Kelly A. Blumenthal,$^{1}$
Joshua E. Barnes,$^{1}$
\\
$^{1}$Institute for Astronomy, University of Hawaii, 2680 Woodlawn Drive, Honolulu, HI 96822, USA}

\date{Accepted 13 June 2018. Received 26 Jun 2017}

\pubyear{2018}

\begin{document}
\label{firstpage}
\pagerange{\pageref{firstpage}--\pageref{lastpage}}
\maketitle

\begin{abstract}
Dynamical interactions between colliding spiral galaxies strongly affect the state and distribution of their interstellar gas. Observations indicate that interactions funnel gas toward the nuclei, fueling bursts of star formation and nuclear activity. To date, most numerical simulations of galaxy mergers have assumed that the gaseous and stellar disks initially have the same distribution and size. However, observations of isolated disk galaxies show that this is seldom the case; in fact, most spirals have as much or more gas beyond their optical radii as they do within. Can gas in such extended disks be efficiently transported to the nuclei during interactions?

To address this question, we examine the effect of various parameters on the transport of gas to the nuclei of interacting galaxies. In addition to the relative radii of the gaseous and stellar disks, these parameters include the pericentric separation, disk orientation, fractional gas mass, presence of a bulge, treatment of gas thermodynamics, and the spatial resolution of the numerical simulation. We found that gas accumulates in most of our simulated nuclei, {but the efficiency of inflow is largely dependent upon the encounter geometry.} Dissipation alone is not enough to produce inflows; an efficient mechanism for extracting angular momentum from the gas is necessary. Several different mechanisms are seen in these experiments. Aside from {mode-}driven inflows {(such as, but not limited to, bars)} and ram-pressure sweeping, both of which have been previously described {and well studied}, we {supply the first quantitative study of an often-seen} process: the formation of massive gas clumps in Jeans-unstable tidal shocks, and their subsequent delivery to the nuclei via dynamical friction. 
 
\end{abstract}


\section{Introduction}
When two galaxies collide, the luminous matter responds to the merging gravitational potentials, which are largely generated by the invisible dark matter halos. Tidal effects manifest in ``bridges'' -- luminous matter bridging the two galaxies -- and ``tails'' -- luminous matter trailing behind one or both of the colliding galaxies. Individual stars within each galaxy have a negligible chance of colliding. However, the galaxies' gas disks are strongly affected by one another; the interaction may trigger galaxy-wide or nuclear bursts of star formation \citep[e.g.,][]{alons2000, barnes2004, evans2008, chien2010} or accretion onto a central supermassive black hole \citep[e.g.,][]{kor2001, springel2005a, hopkins2010, gasp2013, rich2015}. 

\begin{figure}[H]
	\begin{center}
		\includegraphics[width=0.4\textwidth]{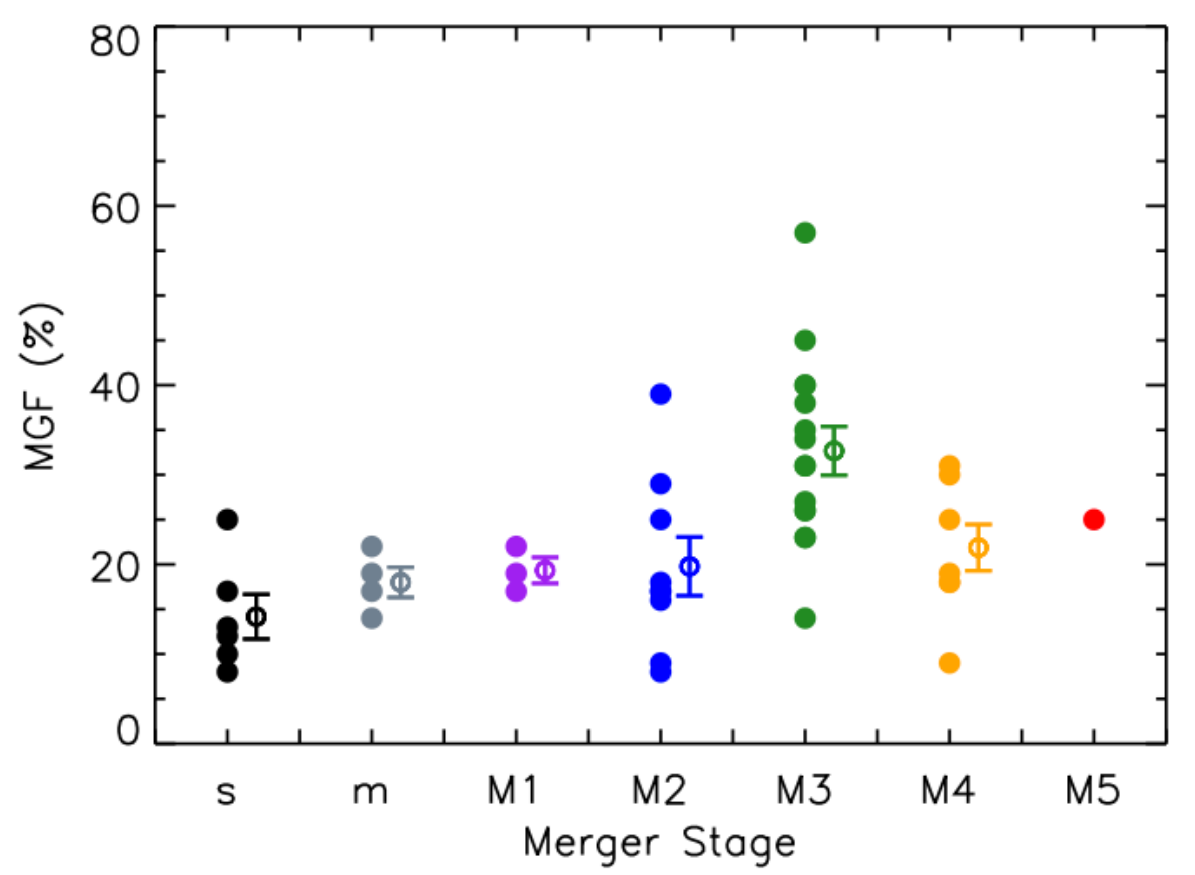}
		\caption{Reproduced from \protect\cite{lar2016}. Molecular gas fraction, $M_{H_{2}} /(M_{*} + M_{H_{2}}$ ), as a function of merger stage: single galaxies (s), minor mergers (m), and major mergers (M$1-$M$5$). At each stage, the mean molecular gas fraction and corresponding uncertainties are shown as filled circles and empty circles with error bars, respectively. The increase in molecular gas content between M1 and M3 may be a direct result of inflows.}
		\label{fig:larson2015}
	\end{center}		
\end{figure}

\begin{figure}
	\begin{center}
		\includegraphics[width=0.4\textwidth]{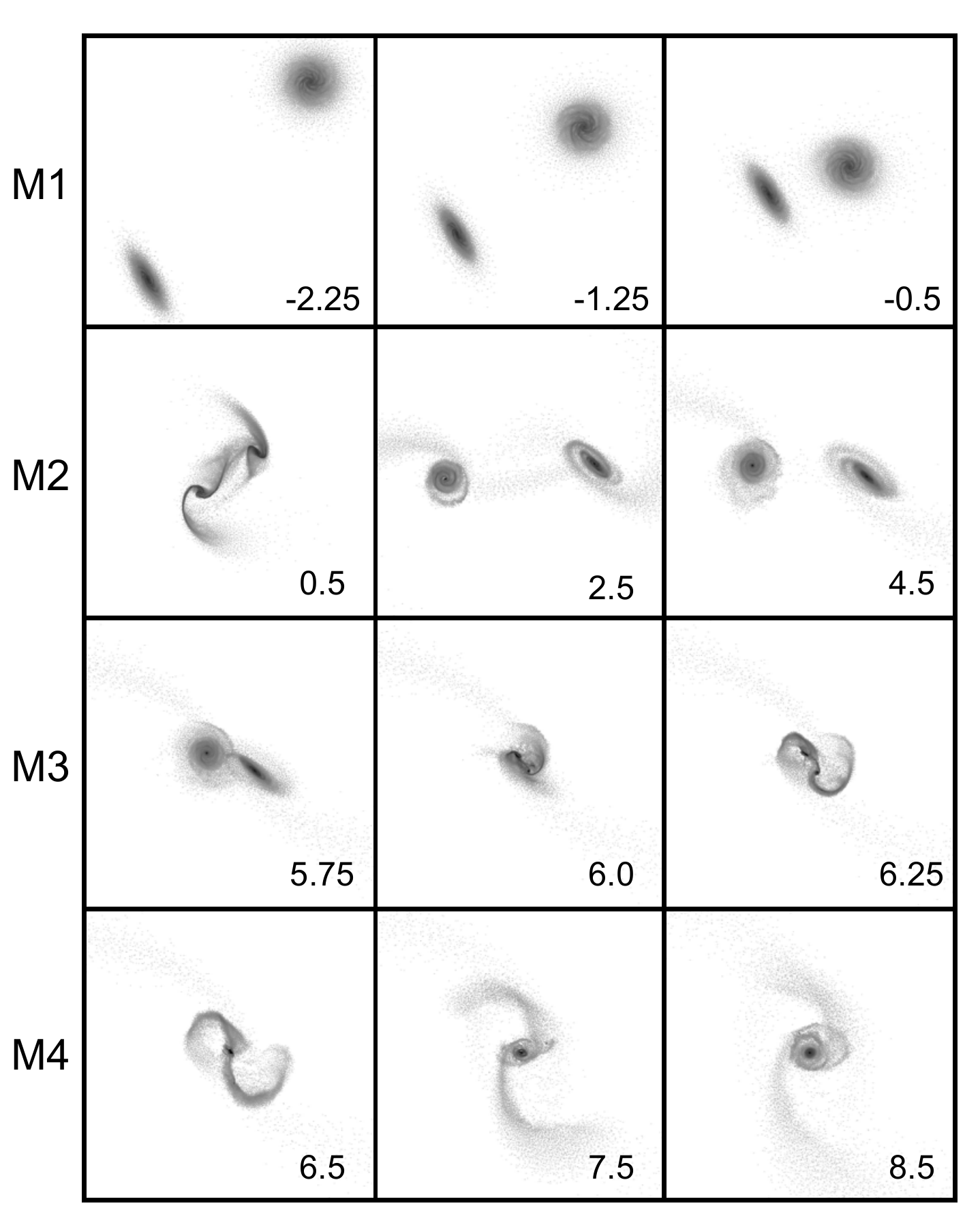}
		\caption{Simulation snapshots showing only the gas particles, organized in the merger stage sequence defined by \protect\cite{lar2016} for M1 through M4. This is encounter B1 (see \S \ref{sec:encounter} for details). Each snapshot is labeled with the time relative to pericenter (in simulation units; see \S \ref{sec:units}). M1: Galaxies are well separated and on their initial approach. M2: Tidal features (bridges and tails) are clearly visible, and prior to second passage. M3: Two individual nuclei are visible in highly disturbed overlapping disks. The tidal tails are still well defined. M4: The two nuclei have now coalesced, but the tidal tails are still visible. 
For full animations of the encounters presented here, please refer to: \url{kelblu.weebly.com/animations.html}}
		\label{fig:mergerstages}
	\end{center}		
\end{figure}

An important tool for probing these changes is the molecular gas: the material from which stars form. \cite{lar2016} devised a merger stage classification scheme that included non-interacting single galaxies (s), minor mergers (m), and major mergers, ranging from before first pericentric passage through final coalescence and post-merger remnant (M$1-$M$5$; see Figures \ref{fig:larson2015} and \ref{fig:mergerstages}). They also derived the molecular gas mass fraction (MGF) as a function of merger classification stage, and found that there appears to be a significant increase in the MGF during M$3-$M$4$, as seen in Figure \ref{fig:larson2015}. Interestingly, the increase in molecular gas content between M1 and M3 corresponds to an increase in the mean IR luminosity. \cite{lar2016} posit that this increase is a direct result of inflows: atomic hydrogen from the outskirts of the galaxy is swept into the central regions and converted to H$_{2}$, which fuels a burst of star formation, and naturally results in an increase of the IR luminosity.

\begin{figure}
	\begin{center}
		\vspace{-12pt}
	    \includegraphics[width=0.4\textwidth]{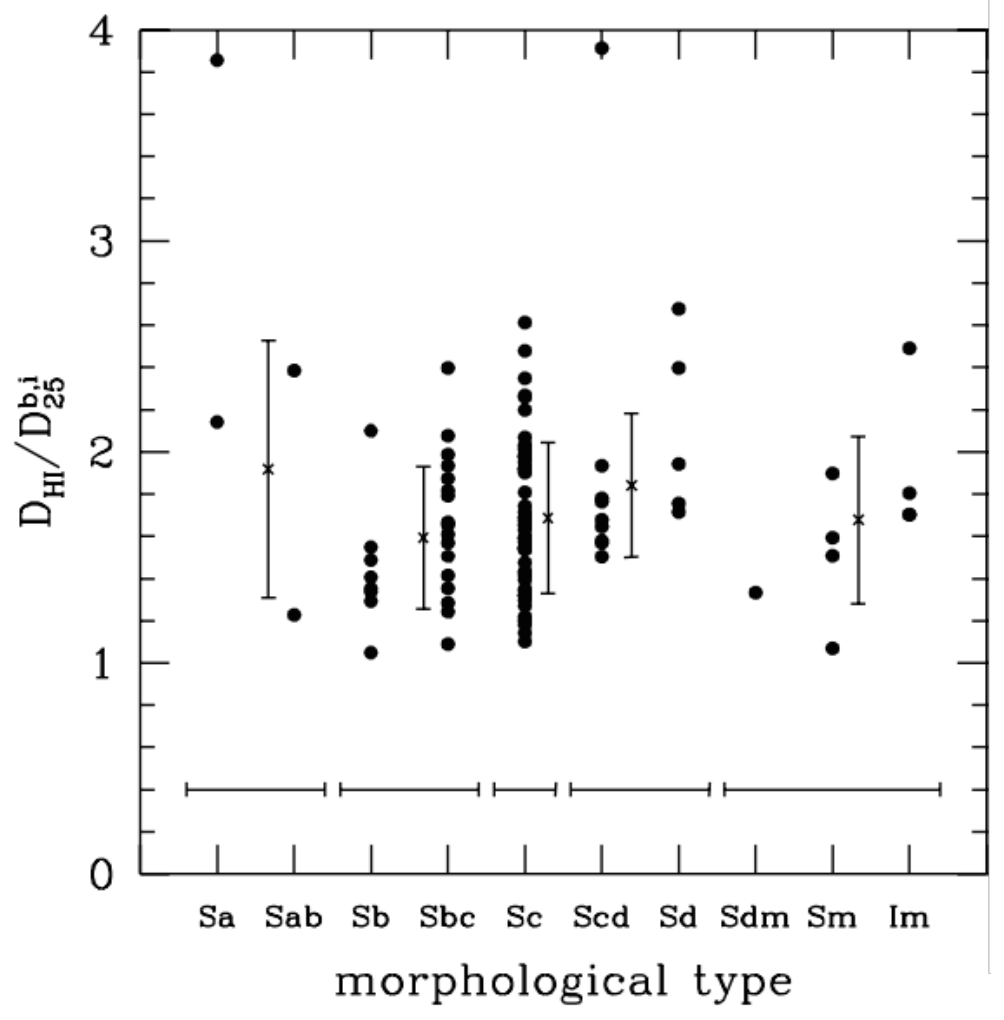}
   		\caption{After \protect\cite{broeils1997}, Figure 3a. This figure shows the ratio of D$_{HI}$ to D$_{25}^{b,i}$, the HI and optical diameters, respectively, as a function of morphology type. In this work, the HI diameter is defined as the point at which the surface density of gas reaches 1 M$_{\odot}$pc$^{-2}$, and the optical diameter is taken as the isophote at 25 mag arcsec$^{2}$. Within each morphology group, the average is represented as an x, with its corresponding 1$\sigma$ dispersion. It is clear that the HI disk is typically larger than the stellar disk.}
		\label{fig:radial}
	\end{center}
\end{figure}

One assumption often made by previous simulations is that the stellar and gaseous disks are initially similar in size. However, using 21cm line observations of about 100 galaxies, \mbox{\cite{broeils1997}} showed that the HI disks of spiral galaxies are always equal to or larger than that of the stellar disks (Figure \ref{fig:radial}). They found that on average there is just as much gas outside the stellar disk as there is inside. 

\vspace{2mm}

This work was motivated by two key questions:
\begin{enumerate}
\item Can the gas in extended gas disks be efficiently transported into the nuclei during interactions?
\item How does the merger-driven inflow mechanism depend on the galaxy structure?
\end{enumerate}

We vary the relative sizes of the gaseous and stellar disks to understand how this parameter affects the nuclear gas fraction. To maintain consistency with observations, we will alter the size of the gas disk within {the range of \cite{broeils1997}} results (that is, $\alpha_{\star} / \alpha_{g}$ = 1 $-$ 2 , where $\alpha_{\star}$ and $\alpha_{g}$ are the inverse scale length of the stellar or gas disks, respectively). Given that the most drastic changes to the state of the gas appear to occur between M2 and M4 (Figure \ref{fig:larson2015}), we will focus only on the time between first and second pericentric passage.

\section{Methods}
\begin{figure}
	\begin{center}
	    \includegraphics[width=0.5\textwidth]{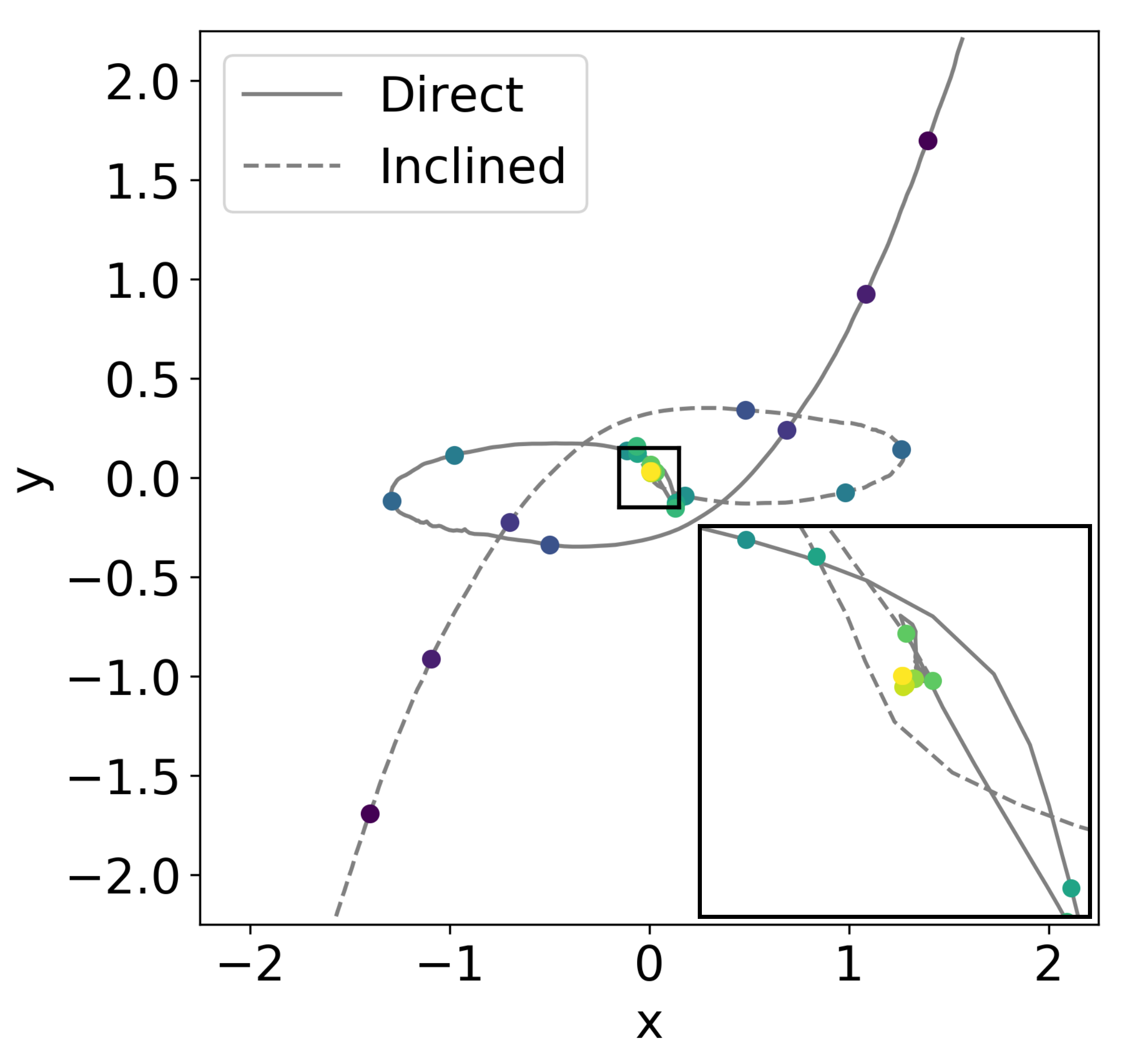}
	    \includegraphics[width=0.5\textwidth]{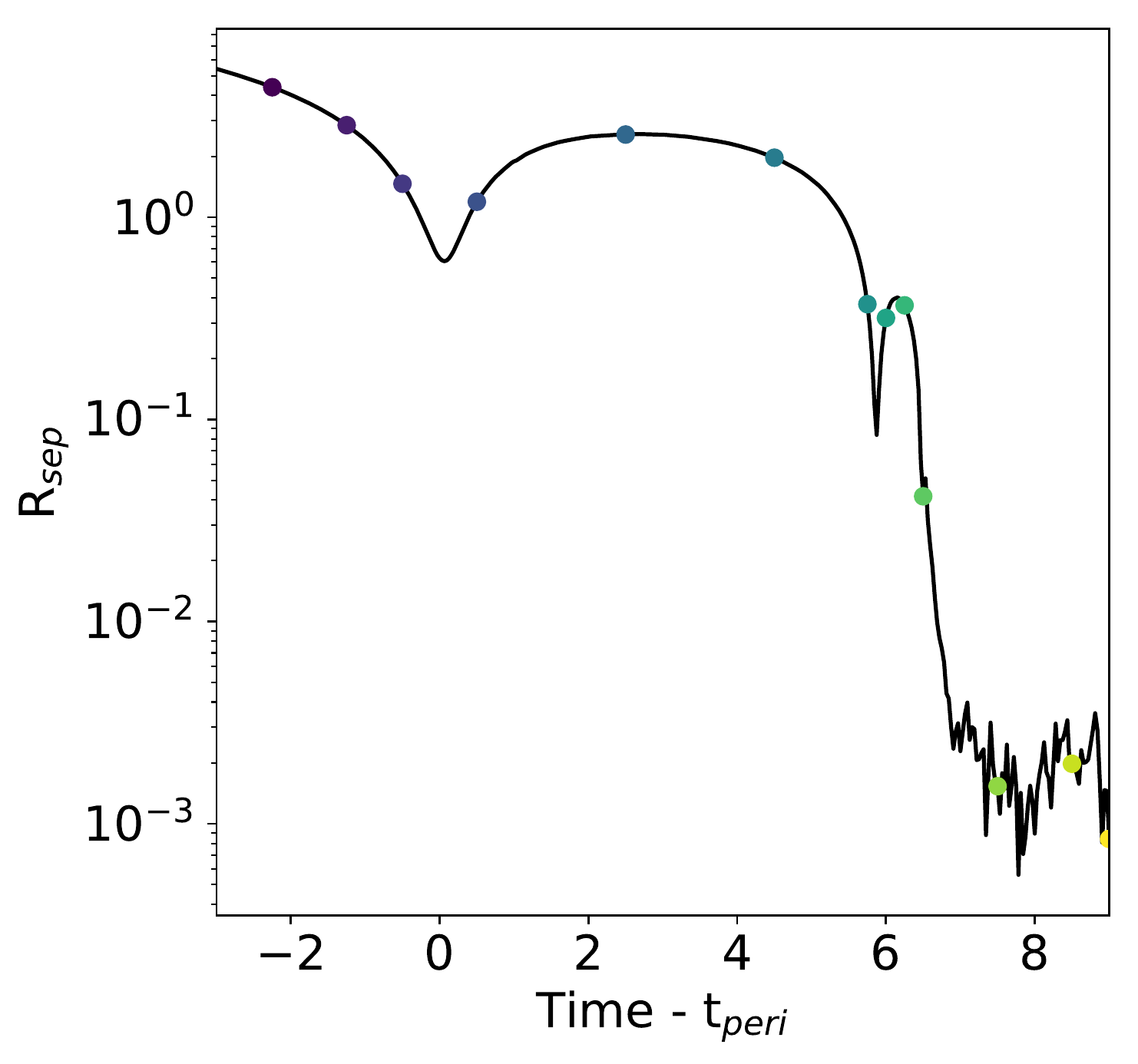}
	    \caption{\textit{Top:} The trajectories of the interacting galaxies in a single encounter simulation. The inset of the top panel shows the details of post-second pericentric passage. Colours indicate a time sequence similar to that of Figure \ref{fig:mergerstages}. \textit{Bottom:} The magnitude of the separation between the two galaxies in the encounter shown in the top panel.}
		\label{fig:orb}
	\end{center}
\end{figure}

In this study, we use N-body/smoothed particle hydrodynamics (SPH) simulations similar to those described in e.g., \cite{barnes2004}. Two identical galaxies are set on initially Keplerian, parabolic ($e=1$) orbits. As the galaxies approach one another, their dark matter halos will undergo gravitational friction, causing their orbits to decay, as seen in Figures \ref{fig:mergerstages} and \ref{fig:orb}. The two galaxies gradually coalesce, eventually forming a single galaxy.  

SPH codes model fluids using smoothed kernels, defined by a kernel smoothing radius, $h$. This radius is adjusted to contain a fixed number of $40$ gas particles, which collectively define the kernel's hydrodynamic properties. This methodology allows for an adaptive resolution; the kernels will be smallest where the fluid density is highest. {For a characteristic density of about $0.015M_{\odot} pc^{-3}$, $h\simeq 120pc$. The smoothing length will scale as $\rho^{-1/3}$, so this value will naturally change depending on disk substructure.} For more details on the SPH code, see Appendix \ref{sec:sphcode}. Energy is conserved to within $1\%$ between first and second pericenter, at which point it fails due to the high densities reached when star formation is not included.

{

It is important to note that this work does not include the effects of star formation, winds, supernovae or feedback due to an active galactic nucleus. The purpose of this study was to examine the simplest scenario possible, in which only gravitational dynamics and hydrodynamics are considered. To that end, we assume an isothermal equation of state (see \S \ref{sec:model} for more details). A good deal of how we currently think we understand fueling comes from decades-old calculations using isothermal gas \citep[e.g.,][]{hernquist1989, barnes1991}. It is thus appropriate to revisit this assumption, but with more accurate galaxy models and better resolution.

This work makes several other assumptions consistent with other studies of idealized simulations. We use a perfectly spherical NFW halo that is truncated at a certain radius, an exponential stellar disk with constant scale height, and a pair of identical galaxies, each of which is initially in near perfect equilibrium. We are exploring the simplest scenario in which only gravitational dynamics and a simplified version of hydrodynamics are considered. In future work, we will assess the validity of the aforementioned assumptions within the context of what is ``observed'' in state-of-the-art cosmological simulations, such as IllustrisTNG \citep[][]{pill2017, wein2017}. In such simulations, we would anticipate the stellar disk to be smaller than the gas disk, as stars are formed from the gas disk based on a density criterion. It is thus important to understand how the relative sizes of stellar and gas disks plays a role in the inflow mechanics of idealized calculations, before we investigate similar questions in much more complex systems. 
}

\subsection{Galaxy Models}
\label{sec:model}
In these models, each galaxy is constructed from four discrete components: a central bulge, a stellar disk, a gas disk, and a dark matter halo. We usually assume a baryon fraction, (M$_{bulge,b}$ + M$_{gas,g}$ + M$_{stars,\star}$)/M$_{total}$, of 10\%, consistent with current predictions of $\Lambda$CDM {\citep[e.g.,][]{planck2013}}.  A full account of the parameters used in our galaxy models is given in Table \ref{table:diskprops}. { We typically assume a gas mass fraction of 25\%, although two cases adopt 12.5\% gas mass (see Table \ref{table:encounters}). Most of our galaxy models are derived from the suite described by \cite{barnes2016}. For comparison with  earlier simulations of galaxy encounters, like those of \cite{barnes1991}, we ran a few simulations with a baryon fraction of 20\%.}

The components of each model are:
\begin{enumerate}
\item \textbf{a bulge} \\
The mass density of the galaxy bulge is modeled according to \cite{jaffe1983},
	\begin{equation}
		\rho(r) = \frac{a_{b} M_{b}}{4\pi r^{2}(a_{b} + r)^{2}}
	\end{equation}
where $M_{b}$ is the total bulge mass and $a_{b}$ is the scale radius of the bulge. 

The Jaffe model is a good representation of a spherical galaxy, and thus serves well to model the galactic bulge. The bulge contains 25\% of the baryons (that is, M$_{b}$/(M$_{b}$ + M$_{g}$ + M$_{\star}$) = 0.25), but is compact with respect to the stellar disk ($a_{b}\alpha_{\star} \approx 0.5$).
\item \textbf{a single-component exponential/isothermal stellar disk} \\
     The stellar disk has an exponential density distribution,
     \begin{equation}
     	\rho(R, z) = \rho_{o} \mathrm{sech}^{2}(z/z_{o})\mathrm{e}^{-R\alpha_{\star}}
     \end{equation}
where, $\Sigma_{o}$ is the central surface density, R is the polar radial coordinate, and $\alpha_{\star}$ is the inverse scale length of the stellar component of the disk. The vertical distribution of the stellar disk is approximated as an isothermal sheet, with constant scale heigh $z_{o}$.
\item \textbf{a single-component exponential gas disk} \\
     The gas is distributed with a similar density distribution as the stellar disk, as shown in Eqn. 2, but with a different scale length, $\alpha_{g}$. {The gas abides by an isothermal (T $\approx 2 \times 10^{4}$ K) equation of state. This temperature is chosen to allow for non-thermal pressure sources (e.g., magnetic fields) \citep[e.g.,][]{barnes2002}.} We assume also that the gas disk is thinner than the stellar disk, and solve the equations of hydrostatic equilibrium to derive its vertical structure.

\item \textbf{a dark matter halo} \\
	The dark matter halo is a \citet*{nfw} model,
	\begin{equation}
	\rho(r) = \frac{a_{h}^{3} \rho_{o}}{r (a_{h} + r)^{2}}
	\end{equation}
where $a_{h}$ is the scale radius of the halo and $\rho_{o}$ is a constant that depends on the halo parameters. This model has $\rho \propto r^{-1}$ as $r \to 0$ and $\rho \propto r^{-3}$ as $r$ tends to large radii. The halo tapers off exponentially at the virial radius \citep[][]{springel1999}.
\end{enumerate}

\begin{figure}
	\begin{center}
	    \includegraphics[width=0.485\textwidth]{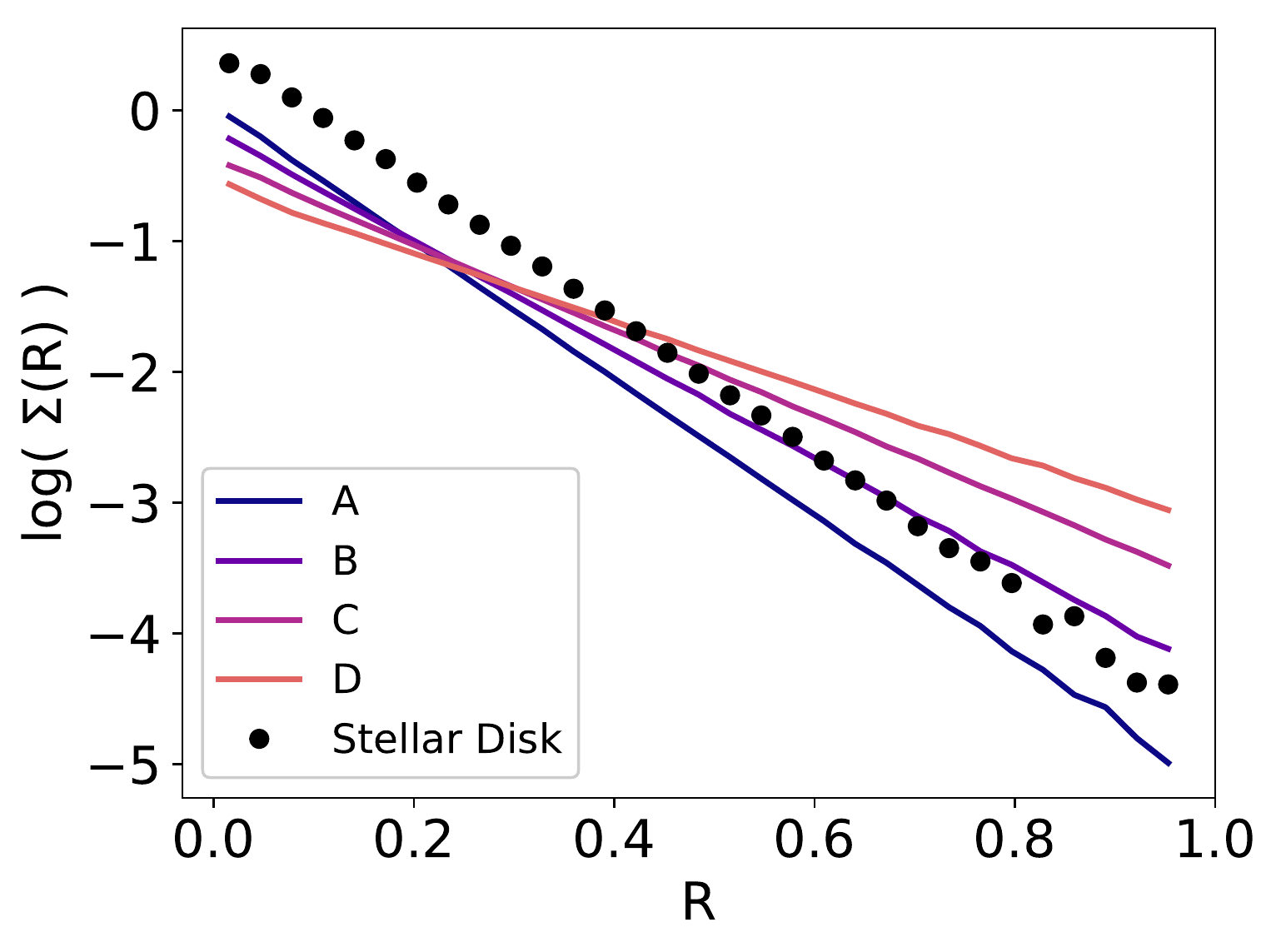}
	    \caption{The initial gas disk surface mass density distributions for each model described in Tables \ref{table:diskprops} and \ref{table:encounters}. Also shown is the stellar disk surface mass density (black dots) for comparison.}
		\label{fig:dists_init}
		\vspace{-12pt}
	\end{center}
\end{figure}

\begin{table*}
\centering
\resizebox{\textwidth}{!}{%
\begin{tabular}{ccccccccccccccc}
\hline
\multirow{3}{*}{ID} & \multirow{3}{*}{$\alpha_{\star} / \alpha_{g}$} & \multicolumn{3}{c}{\multirow{2}{*}{\textbf{Bulge}}} & \multicolumn{6}{c}{\textbf{Disk}}                                                             & \multicolumn{3}{c}{\multirow{2}{*}{\textbf{Halo}}} & \multirow{3}{*}{$\epsilon$} \\
                    &                                                & \multicolumn{3}{c}{}                                & \multicolumn{3}{c}{Gas}                        & \multicolumn{3}{c}{Stars}                    & \multicolumn{3}{c}{}                               &                             \\ \cline{6-11}
                    &                                                & M$_{b}$         & a$_{b}$         & N$_{b}$         & M$_{g}$  & $\alpha_{g}$              & N$_{g}$ & M$_{\star}$ & $\alpha_{\star}$ & N$_{\star}$ & M$_{h}$         & a$_{h}$         & N$_{h}$        &                             \\ \hline
A-D                 & {[}1.0, 1.25, 1.6, 2.0{]}                      & 0.0625          & 0.04            & 16384           & 0.046875 & {[}12.0, 9.6, 7.5, 6.0{]} & 49152   & 0.140625    & 12.0             & 36864       & 2.25            & 0.25            & 147456         & 0.0025                      \\ \hline \hline
E & 1.0 & 0.0625 & 0.04 & 16384 & 0.046875 & 12.0 & 49152 & 0.140625 & 12.0 & 36864 & 1.0 & 0.25 & 65536 & 0.0075\\ \hline
E' & 1.0 & 0.0625 & 0.04 & 16384 & 0.0234375 & 12.0 & 24576 & 0.1640625 & 12.0 & 43008 & 1.0 & 0.25 & 65536 & 0.0075\\ \hline
\end{tabular}
}
\caption{This table provides the specific parameters, in simulations units, used for each of the disks modeled here. M$_{b,g,\star,h}$ are the bulge, gas disk, stellar disk and halo masses, respectively. $a_{b,h}$ are the bulge and halo length scales, while $\alpha_{g,\star}$ are the gas and stellar disk inverse length scales. N$_{b,g,\star,h}$ are the number of particles used for each component of the model. The smoothing length, $\epsilon$, is also given. Note that the gas disk size increases with $\alpha_{\star}/\alpha_{g}$.}
\label{table:diskprops}
\end{table*}

\subsubsection{Simulation Units}
\label{sec:units}
All simulated quantities quoted in this study are in units where the gravitational constant, $G = 1$. That is, we set
\begin{equation}
\label{eqn:gravconst}
	G = 6.67 \times 10^{-11} kg^{-1} m^{3} s^{-2} = 1 \mathcal{M}^{-1} \mathcal{L}^{3} \mathcal{T}^{-2}
\end{equation}
to constrain simulation units of mass, length and time: $\mathcal{M, L}$ and $\mathcal{T}$. Another condition, based on the specific internal energy of the gas, fixes the velocity scale
\begin{equation}
\label{eqn:five}
	u_{int} = \frac{3}{2} \left( \frac{k_{B} T}{\mu m_{H}} \right) \simeq 2.0 \times 10^{8} m^{2} s^{-2} = 0.014 \mathcal{L}^{2} \mathcal{T}^{-2}
\end{equation}
where $k_{B}$ is the Boltzmann constant, $\mu$ is the molecular mass, $T$ is the temperature of the interstellar medium (ISM), and 0.014 is the value for $u_{int}$ chosen in simulation units. Note that the constant $\frac{3}{2}$ in this equation is appropriate for a monatomic gas. We also assume that the ISM is comprised of 75\% H and 25\% He by mass (for this composition, and assuming a totally unionized ISM, $\mu \simeq 1.23$). If we choose a scale for the length, then the time unit follows from Equation \ref{eqn:five}. Using those two values, we can easily derive a mass unit from Equation \ref{eqn:gravconst}. For example, if we take $\mathcal{L}$ = 30 kpc, then $\mathcal{T}$ = 2.5 $\times 10^{8}$ yrs, and $\mathcal{M} = 1 \times 10^{11}$ M$_{\odot}$.

\subsubsection{Galaxy Mass Models}

Each galaxy mass model is assigned a letter (A through E). The first four models differ only in the relative size of their gas and stellar disks: the stellar and gas disks of model A are the same size, while the gas disk scale length of the D model is twice that of its stellar disk. Models B and C lie within this range. This is illustrated in Figure \ref{fig:dists_init}, which shows the initial gas surface density distributions for each of the galaxy models. The ratio of the gas disk size to the stellar disk size (i.e., $\alpha_{\star}/\alpha_{g}$) ranges from 1.0 to 2.0. 

After completing several encounters using mass models A-D, we found that none of our simulations formed bars. In an attempt to reproduce bars as seen in earlier simulations \citep[e.g.,][]{barnes1991, springel2005b, hay2014}, we ran two ``legacy models.'' These E models are similar to those of \cite{barnes1991}, which were known to form bars.  The prime (i.e., E') in Table \ref{table:encounters} indicates a lower gas fraction.

We used Plummer smoothing when computing the gravitational potential \citep[e.g.,][]{aarseth1963, barnes2012}. That is, the cannonical $r^{-1}$ potential becomes $(r^{2} + \epsilon^{2})^{-1/2}$, where $\epsilon$ is the smoothing parameter. {Typically, the gravitational smoothing parameter is $\epsilon = 0.0025$ (or, roughly 75 pc). The legacy models use a larger smoothing length: $\epsilon = 0.0075$ (or, roughly 225 pc). These values are the same for all particle types, and are also listed in Table \ref{table:diskprops}. The legacy models also contain a less massive halo.}

As a {test of} our galaxy models, we allowed four isolated galaxies (one for each of the original mass models, A-D) to evolve for the entire encounter time. We measured the amount of gas within a defined nuclear radius (details in \S \ref{sec:analysis}) and found that little material flows into the nucleus. The inflow rate is nearly constant, with about 0.15\% to 0.375\% of the gas reaching the centre per unit of simulation time {(see the bottom panel of Figure \ref{fig:inflow_retro} for the inflow of both the A and D isolated control galaxies).} This slow inflow is due to weak torques provided by transient spiral patterns and artificial viscosity. {We see no clump formation in these control models, which indicates that the clumps found in our encounters are indeed a byproduct of galaxy interactions.}

\subsection{Encounter Models}
\label{sec:encounter}
The encounters in this work are described in Table \ref{table:encounters}. To better understand how the ratio of gaseous and stellar disk scales affect dynamical inflows, we tested the effects of several other parameters. These include (1) the disk orientation, (2) the pericentric separation, (3) the fractional gas mass, and (4) the spatial resolution of the simulation. 

In Table \ref{table:encounters}, the ID letters from Table \ref{table:diskprops} are given modifiers to describe the interaction in more detail. The numbers tell us about the pericentric passage: wide (1; r$_{p}$ = 0.5), intermediate (2; r$_{p}$ = 0.2 or 3; r$_{p}$ = 0.25) or close (4; r$_{p}$ = 0.125). 

Previous studies, from \cite{tt1972} on, have shown that prograde encounters of two galaxies in the same orbital plane elicit the strongest response from the disks, forming pronounced tidal features. To the extent that the dynamics of the two disks are largely decoupled at early times, we can explore the effects due to collision geometry on the outcome of the interaction by studying an encounter with one in-plane galaxy (inclination, $i = 0^{\circ}$) and an inclined galaxy (in this case, $i = 72^{\circ}$). Similarly, we can do the same for retrograde disks ($i = 180 ^{\circ}$) and inclined retrograde disks ($i = 109^{\circ}$). The subscript $r$ denotes a retrograde encounter.

\begin{table}
\centering
\begin{tabular}{l|c|l|c|c|c|c|c|c|}
\hline Encounter & r$_{p}$ & $\alpha_{\star}/\alpha_{g}$ & $i_{1}$$^{\circ}$ & $i_{2}$$^{\circ}$ & $f_{gas}$\\ \hline
A1 & 0.5 & 1.0 & 0 & 72 & 0.25 \\ 
A1$_{r}$ & " & 1.0 & 180 & 109  & " \\
B1 & " & 1.25 & 0 & 72  & " \\
C1 & " & 1.6 & 0 & 72  & " \\
D1 & " & 2.0 & 0 & 72 & " \\ 
D1$_{r}$ & " & 2.0 & 180 & 109 & "\\ \hline 
A2 & 0.25 & 1.0 & 0 & 72 & 0.25\\ 
D2 & " & 2.0 & 0 & 72 & "\\ \hline
E3 & 0.2 & 1.0 & 0 & 72 & 0.25\\
E3' & " & 1.0 & 0 & 72 & 0.125\\ \hline
A4 & 0.125 & 1.0 & 0 & 72 & 0.25\\
D4 & " & 2.0 & 0 & 72 & "\\ \hline
\end{tabular}
\caption{The above table summarizes the encounters studied. Here, r$_{p}$ is the pericentric separation, $i_{1, 2}$ are the two galaxies' inclinations with respect to the orbital plane in degrees, and $f_{gas}$ is the fractional amount of gas defined as f$_{gas} = M_{gas} / (M_\star + M_{gas})$.}
\label{table:encounters}
\end{table}

\section{Analysis}
\label{sec:analysis}
When two galaxies collide, their halos spiral toward one another and eventually merge to form a single galaxy. Tidal and hydrodynamic effects disrupt the stellar and gas disks, producing torques which drive material into the galaxies' nuclei. In most of the encounters listed in Table \ref{table:encounters}, we observe substantial amounts of gas flowing into the centre of each galaxy between first and second pericentric passage. In the following sections, we will look at the inflow of material over time in select cases, and then discuss the different mechanisms driving that inflow.

In order to accurately track the inflows, we must be careful of how we define the nucleus of a galaxy. {Here, we chose} the inner $\alpha_{\star} R = 0.15$, or roughly the central kiloparsec in diameter, {which is} consistent with observational definitions \citep[e.g.,][]{que2016}. {Visual inspection showed that this choice was reasonable throughout the interaction history for all collision realizations.}

\subsection{Inflow}
\label{sec:inflow}
\begin{figure}
	\begin{center}
		\includegraphics[width=0.485\textwidth]{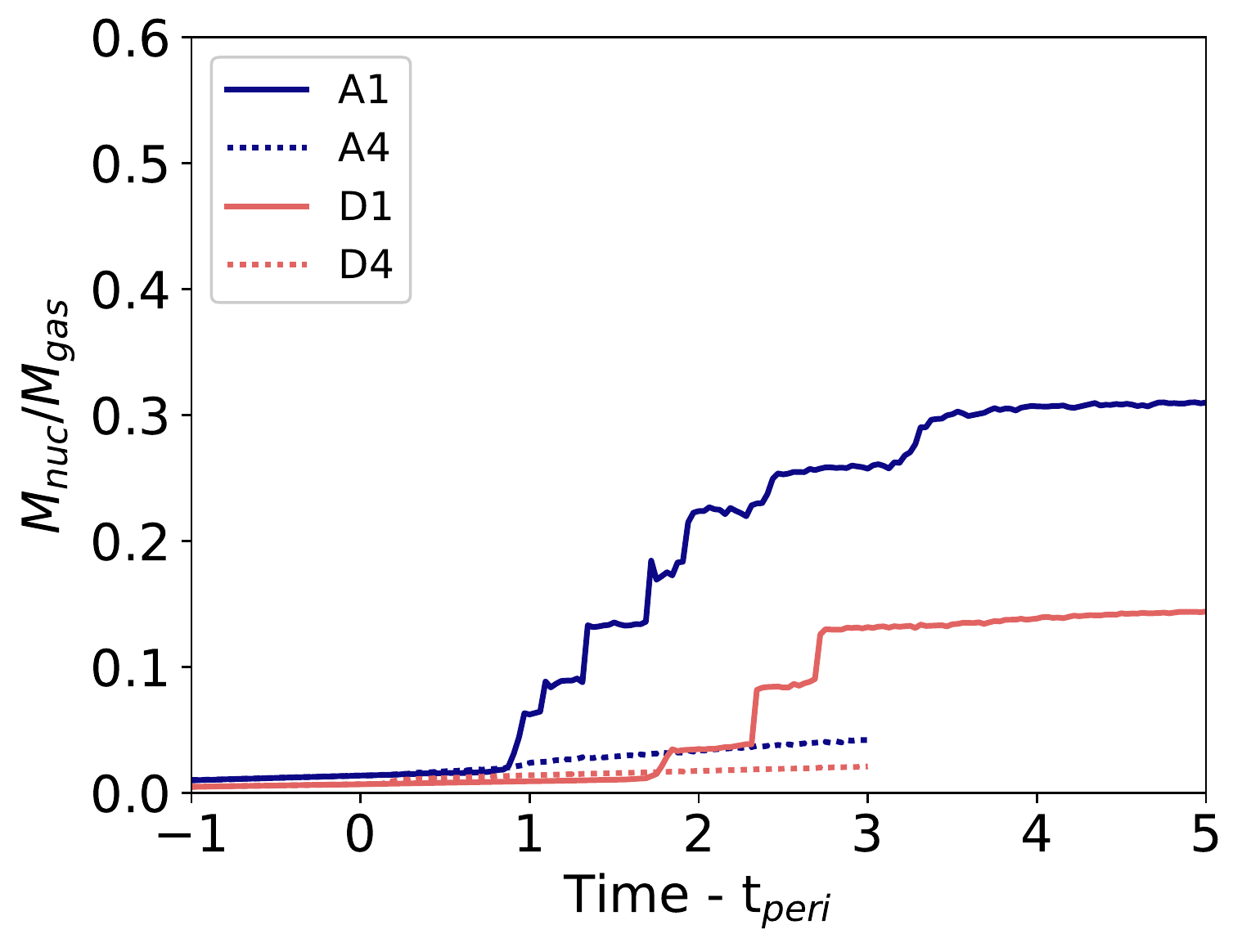}
	    \includegraphics[width=0.485\textwidth]{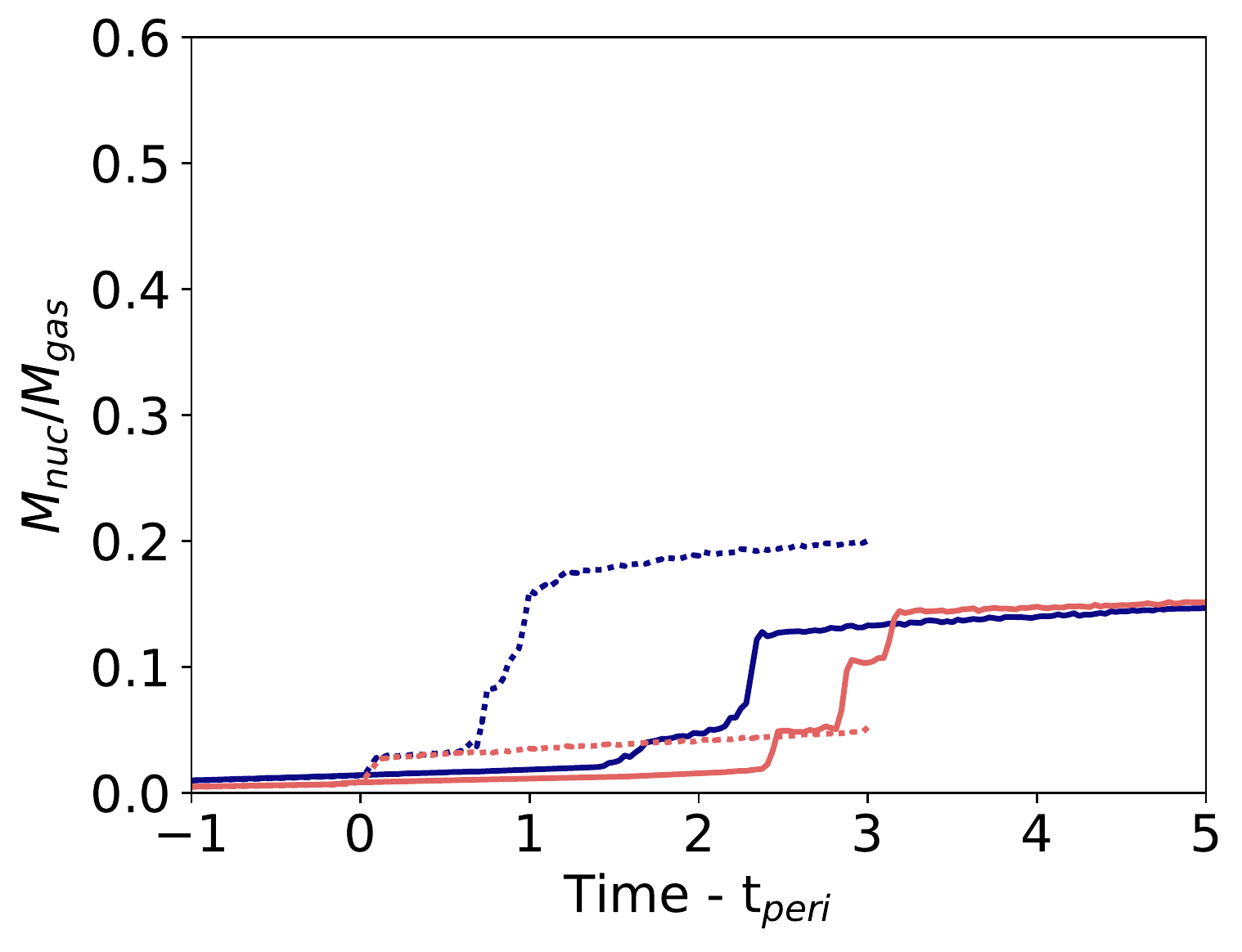}
	    \caption{The fractional amount of gas in the nuclear region is plotted as a function of time for the A1, A4, D1 and D4 models, both the direct (top) and inclined (bottom) disks.}
		\label{fig:inflow_d1}
	\end{center}
\end{figure}

Figure \ref{fig:inflow_d1} shows the fractional amount of gas within {the nucleus} as a function of time for the direct disk (top) and inclined disk (bottom) of the encounter models A1, A4, D1, and D4. The times shown are well before first pericenter until just before second pericenter. These encounters are selected to compare the gas inflows for extreme values of the $\alpha_{\star}/\alpha_{g}$ ratio and pericentric separation. Sharp increases in nuclear gas fraction reflect the arrival of massive clumps of material. 

Inflow is fastest in the $\alpha_{\star}/\alpha_{g} = 1.0$ disks, with large amounts of gas accumulating regardless of pericentric distance or disk inclination. Hereafter we consider encounter D1 as the ``canonical'' encounter, because observations (e.g., Figure \ref{fig:radial}) indicate that gas disks are on average nearly twice as extended as stellar disks. The D1 encounter produces somewhat more modest inflow in both the direct and inclined disks, but much later: several time units after pericentric passage. The direct and inclined disks of B1 and C1 behave as one might expect from the bounds created by A1 and D1, showing a clear correlation of nuclear inflow with decreasing size gas disk. Remarkably, there is virtually no inflow in the direct D4 encounter, despite the violent tidal interaction of the galaxies. 

\begin{figure}
	\begin{center}
		\includegraphics[width=0.485\textwidth]{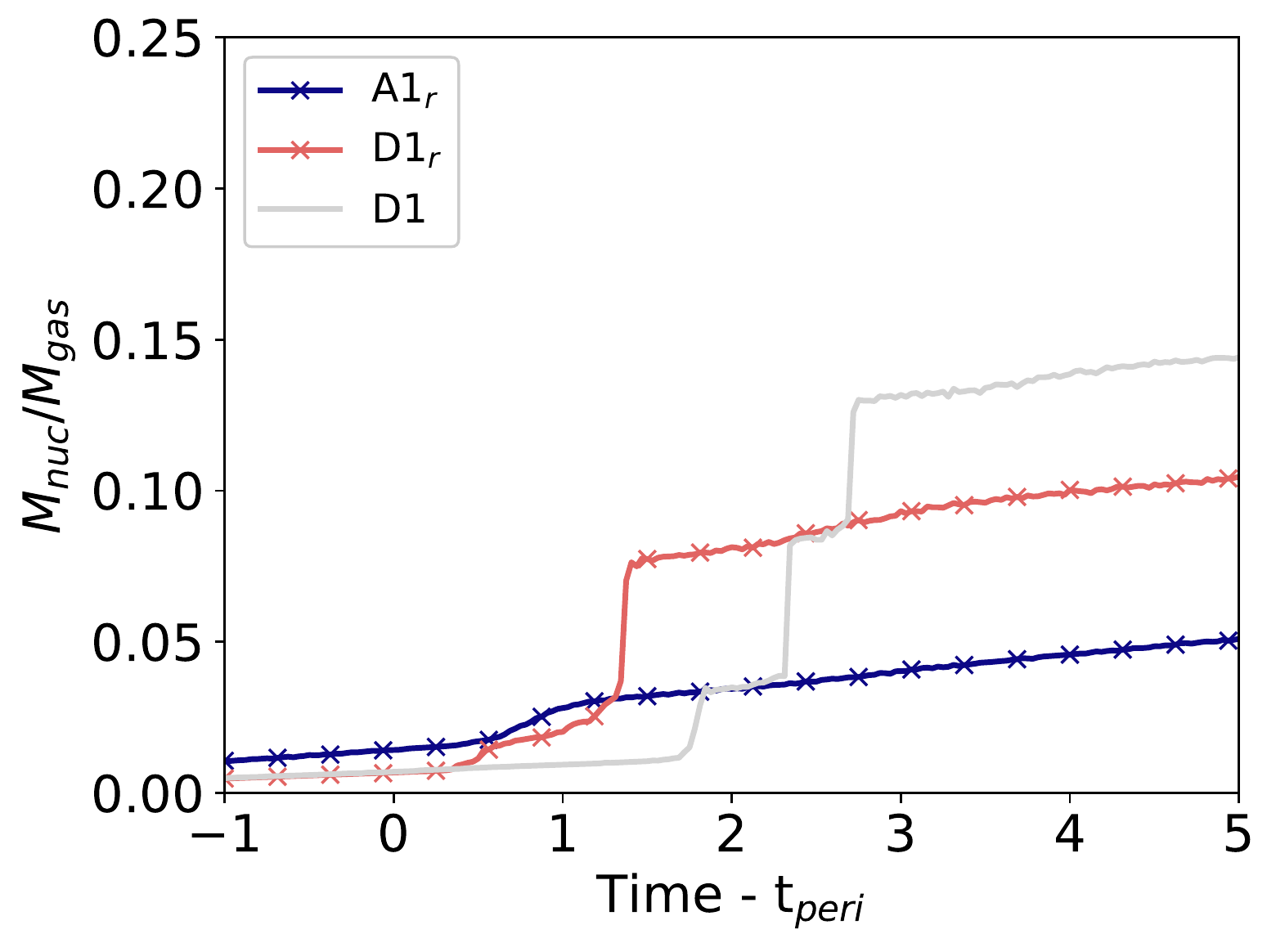}
	    \includegraphics[width=0.485\textwidth]{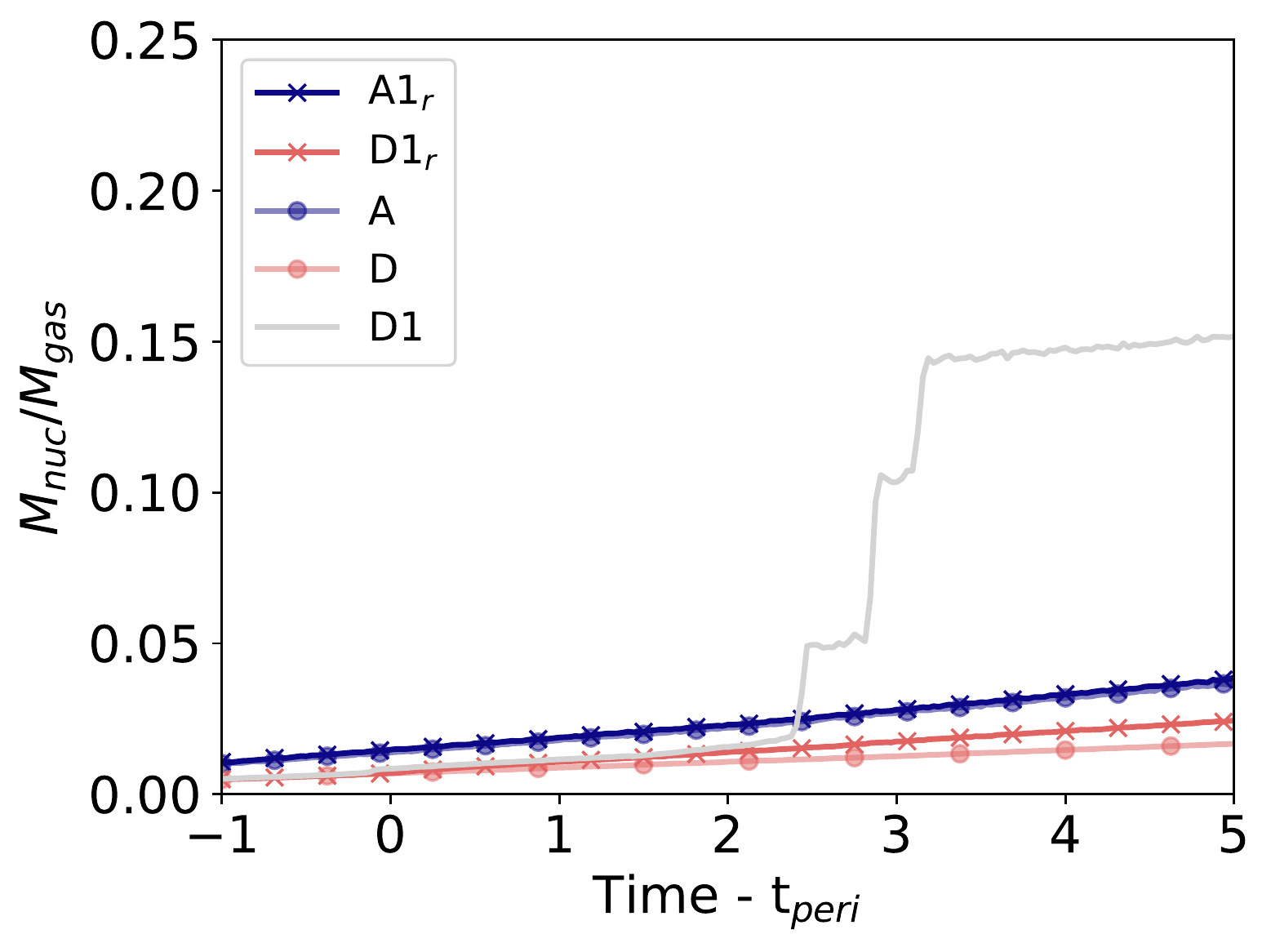}
	    \caption{Here, the nuclear inflow is shown for the A1$_{r}$, and D1$_{r}$ models. The retrograde disk is shown on the top, while the retrograde inclined disk is shown in the bottom panel. The D1 encounter model is also shown for scale to Figure \ref{fig:inflow_d1}. Also shown in the bottom panel is the inflow for the A and D isolated galaxies. The inflow seen in the inclined retrograde passages is very nearly what one might expect from an isolated galaxy.}
		\label{fig:inflow_retro}
	\end{center}
\end{figure}

\begin{figure}
	\begin{center}
	    \includegraphics[width=0.485\textwidth]{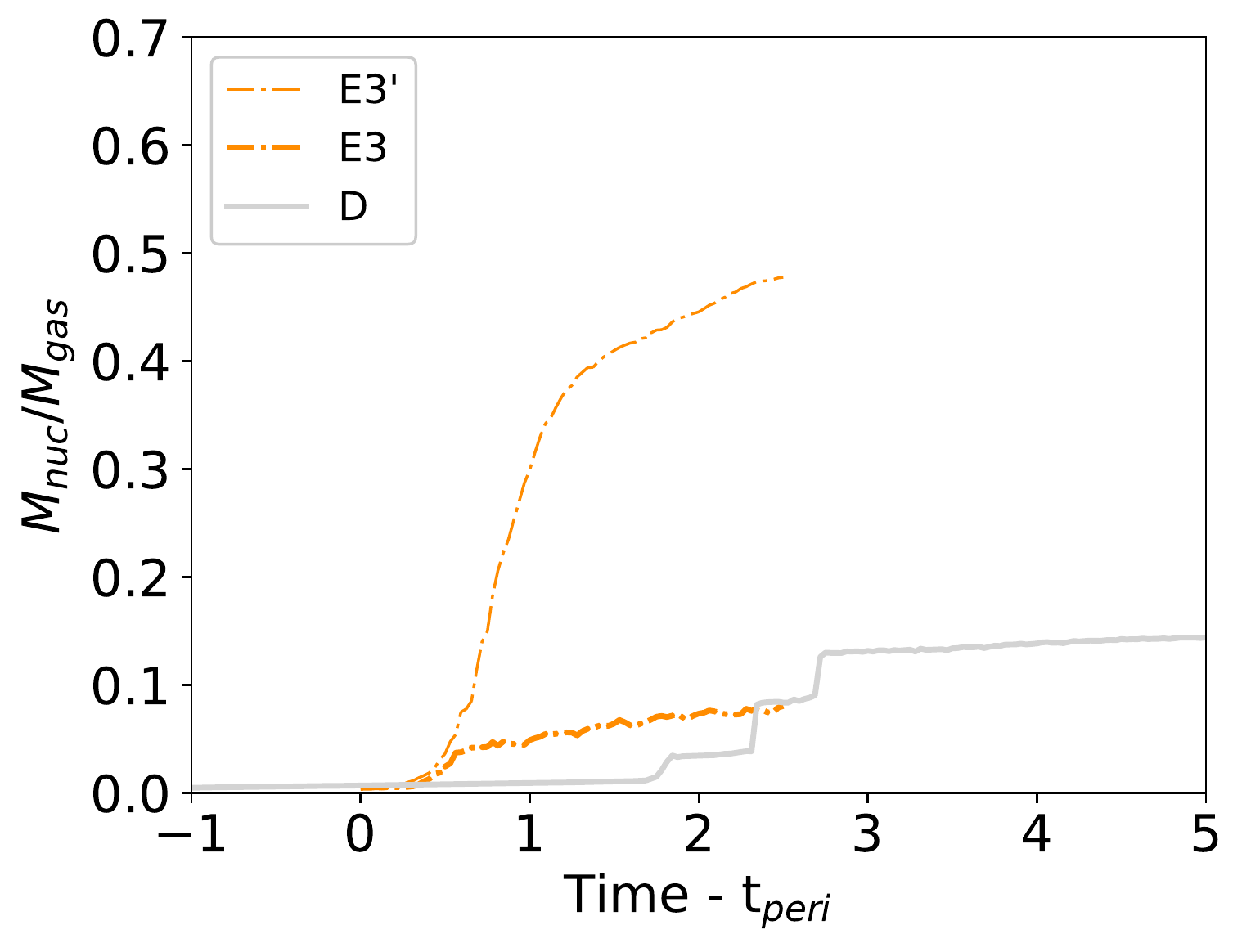}
	    \includegraphics[width=0.485\textwidth]{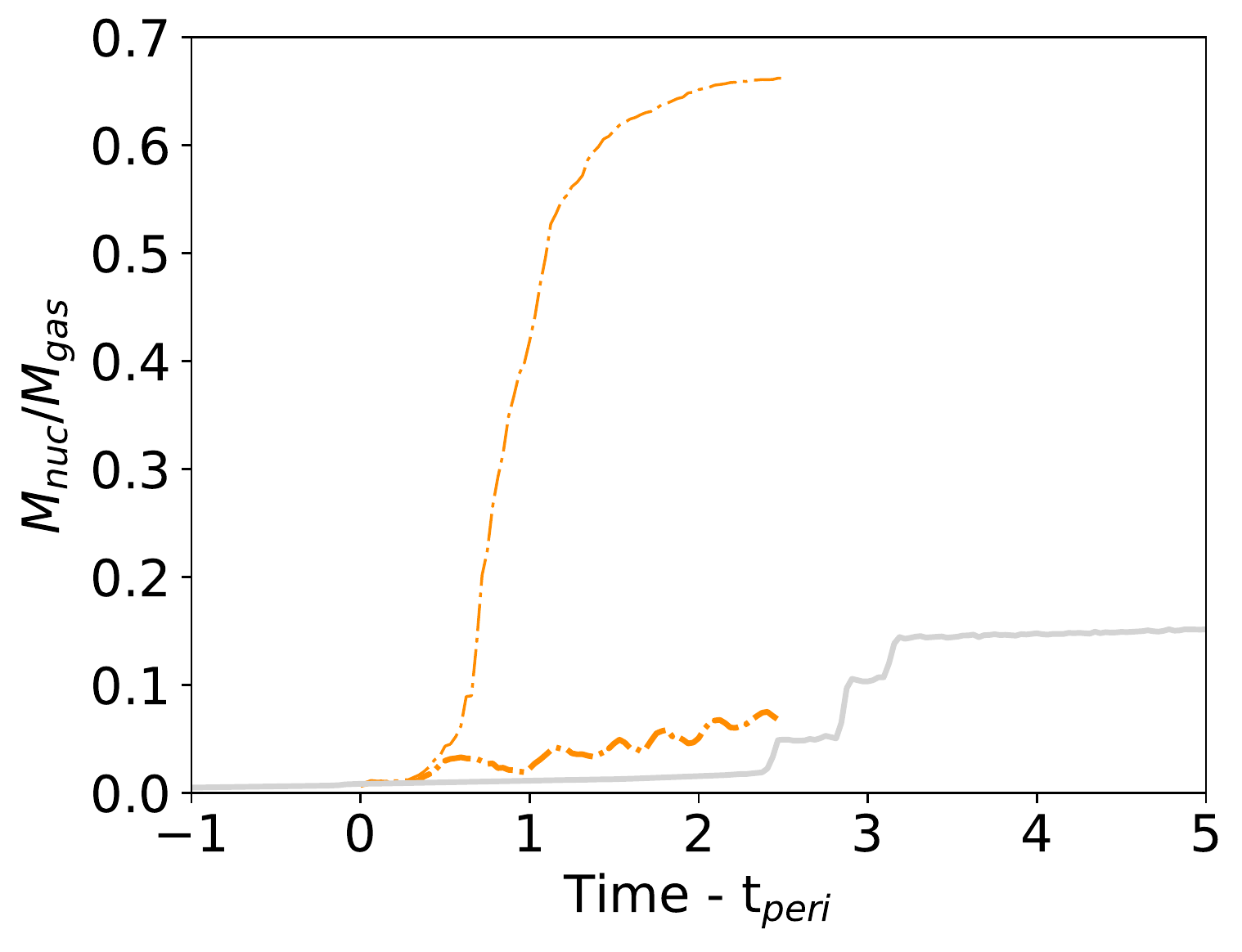}
	    \caption{Here, the nuclear inflow is shown for the legacy E3 and E3' models. The top panel shows the inflow for the direct disk, and the bottom panel shows the same for the inclined disk. The D1 encounter model is also shown for scale to Figure \ref{fig:inflow_d1}.}
		\label{fig:inflow_d2}
	\end{center}
\end{figure}

In a similar vein, the inclined retrograde disks ($i=109^{\circ}$) also show almost no inflow (Figure \ref{fig:inflow_retro}). The gradual inflows seen in the inclined retrograde disks are likely due to the same process driving inflow in the isolated galaxies: torques associated with the spiral pattern slowly drive material inward. The bottom panel of Figure \ref{fig:inflow_retro} compares the relevant isolated disks {(also our control sample)} to the inflow experienced by the inclined retrograde disks. These gradual increases in the nuclear gas fraction seen in Figure \ref{fig:inflow_retro} are visible largely because the vertical scale is significantly {smaller compared to Figures \ref{fig:inflow_d1} and \ref{fig:inflow_d2}}; modest upward trending plateaus can also be seen in other figures (e.g., in the bottom panel of Figure \ref{fig:inflow_d1}, for $t - t_{peri}$ between 3 and 5). 

When inflow does occur, however, the retrograde passages behave very differently than the prograde encounters. Here, extended gas disks are \textit{more} effective at producing inflows (see the top panel of Figure \ref{fig:inflow_retro}). The retrograde disks appear also to react to the encounter faster, likely due to the strong hydrodynamic interaction that occurs near pericenter \citep[as seen in e.g.,][]{capelo2017}.

Comparing the top and bottom panels of Figures \ref{fig:inflow_d1} and \ref{fig:inflow_retro}, we can easily note the complex relationship between inflow and orbital geometry (e.g., inclination and pericentric separation). Closer passages reduce inflow in the direct disk and increase inflow in the inclined disks. {This trend is generally followed by intermediate pericentric passages. A strong tidal interaction does not necessarily imply a large inflow; hydrodynamic forces counter the tidal reaction of an interaction with varying success, depending on the properties of the encounter.} 

Figure \ref{fig:inflow_d2} shows the nuclear inflow for the legacy models (recall, these models are those which have a larger smoothing length than the models previously discussed). These produce prompt and substantial inflows that are nearly 2-3 times larger than the modern models. Comparing the inflow of E3' to that of D1 (shown in grey), we note that the curve is much more smooth, indicating that clumps are not driving inflow in this case. These inflows are remarkably sensitive to gas content; the encounter with 12.5\% gas drives a larger absolute amount of gas inward than does its gas-rich counterpart. The bottom panel of Figure \ref{fig:inflow_d2} shows that the inclined disk of the 12.5\% gas legacy model is also more effective at producing inflow, while the inflow in the 25\% gas legacy model does not appear to be as sensitive to geometry.

As the cases described above make clear, the accumulated nuclear gas has a complex relationship with encounter parameters such as inclination angle and pericentric separation. \cite{tt1972} showed that the tidal response is strongest in a direct prograde disk. We might naively assume that inflow is strongest in the encounters with the strongest tidal response \citep[e.g., direct and close encounters; ][]{barnes1996, mihos1996}. We have shown here that this is not necessarily the case. In fact, the rate and magnitude of the inflow exhibits a complex and somewhat counter-intuitive dependence the circumstances of an encounter. {This trend is also seen in the studies of e.g., \cite{renaud2014} and \cite{dimatteo2007}.} This suggests that there must be a competing force which compensates for a weak tidal response, thereby drawing more material to the nucleus. Below we will show that the hydrodynamic force plays a key role. 

\begin{figure}
	\begin{center}
	    \includegraphics[ width=0.485\textwidth]{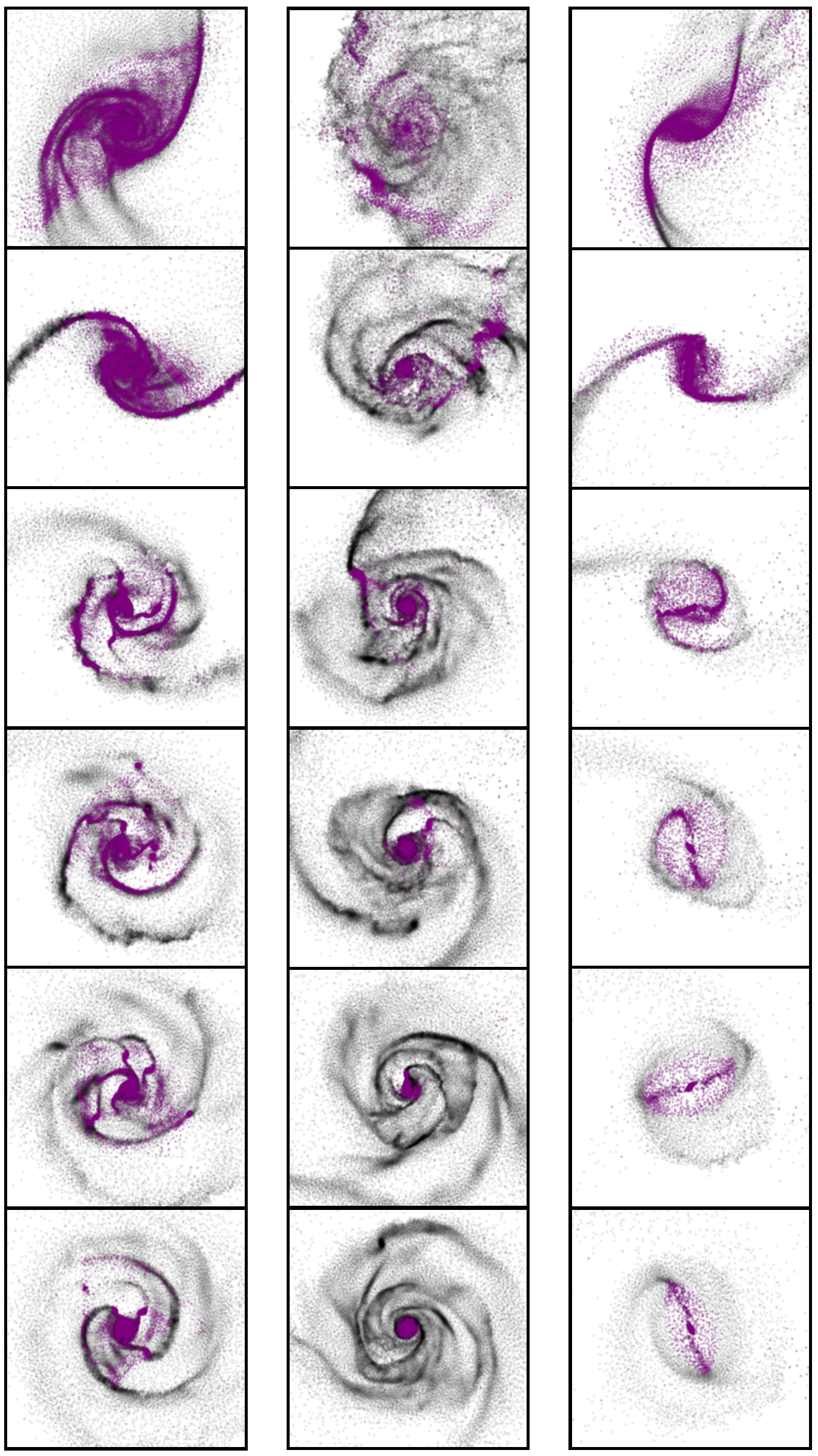}
	    \caption{\textit{Left:} Clump-induced inflow (encounter A1) gradually delivers material to the nucleus. \textit{Middle:} Ram-pressure sweeping (encounter D1$_{r}$) dominates only in retrograde disks. \textit{Right:} {Mode}-driven inflow (encounter E3') rapidly drives gaseous material into the galaxy's centre. Plotted in purple are gas particles in the Lagrangian volume. The remaining gas is displayed in greyscale. All snapshots are {0.6x0.6} length units in size and represent evenly spaced time intervals between $t - t_{peri} = 0.25 - 1.5$. Note that the number of purple particles plotted remains the same in each snapshot of a given encounter. The particles within this Lagrangian volume become much more centrally concentrated as time goes on so they become visually less dominant.}
		\label{fig:bars_clumps}
	\end{center}
\end{figure}

\subsection{Inflow Mechanisms}
The inflows just described are driven by three different mechanisms: clump-driven inflow, ram-pressure sweeping, and {mode-}driven inflow. Figure \ref{fig:bars_clumps} illustrates all three mechanisms, with clump-driven, ram pressure, and {mode-}driven inflow on the left, middle and right panels, respectively. 

In order for inflow to occur, the gas must lose angular momentum via torques which we can calculate directly from the particle configuration. To study these torques, we defined a Lagrangian volume of particles (as in \cite{barnes1991}) which lie within the nucleus (R $\le$ 0.0125, shown in purple in Figure \ref{fig:bars_clumps}) at the time, prior to second pericentre, when the inflow curves end in Figures \ref{fig:inflow_d1} - \ref{fig:inflow_d2}. This ensures that we have allowed sufficient time, after the first interaction, to maximize the mass of gas in the nucleus prior to second encounter. 

\subsubsection{Clump-Driven Inflow}
\begin{figure}
	\begin{center}
	    \includegraphics[width=0.485\textwidth]{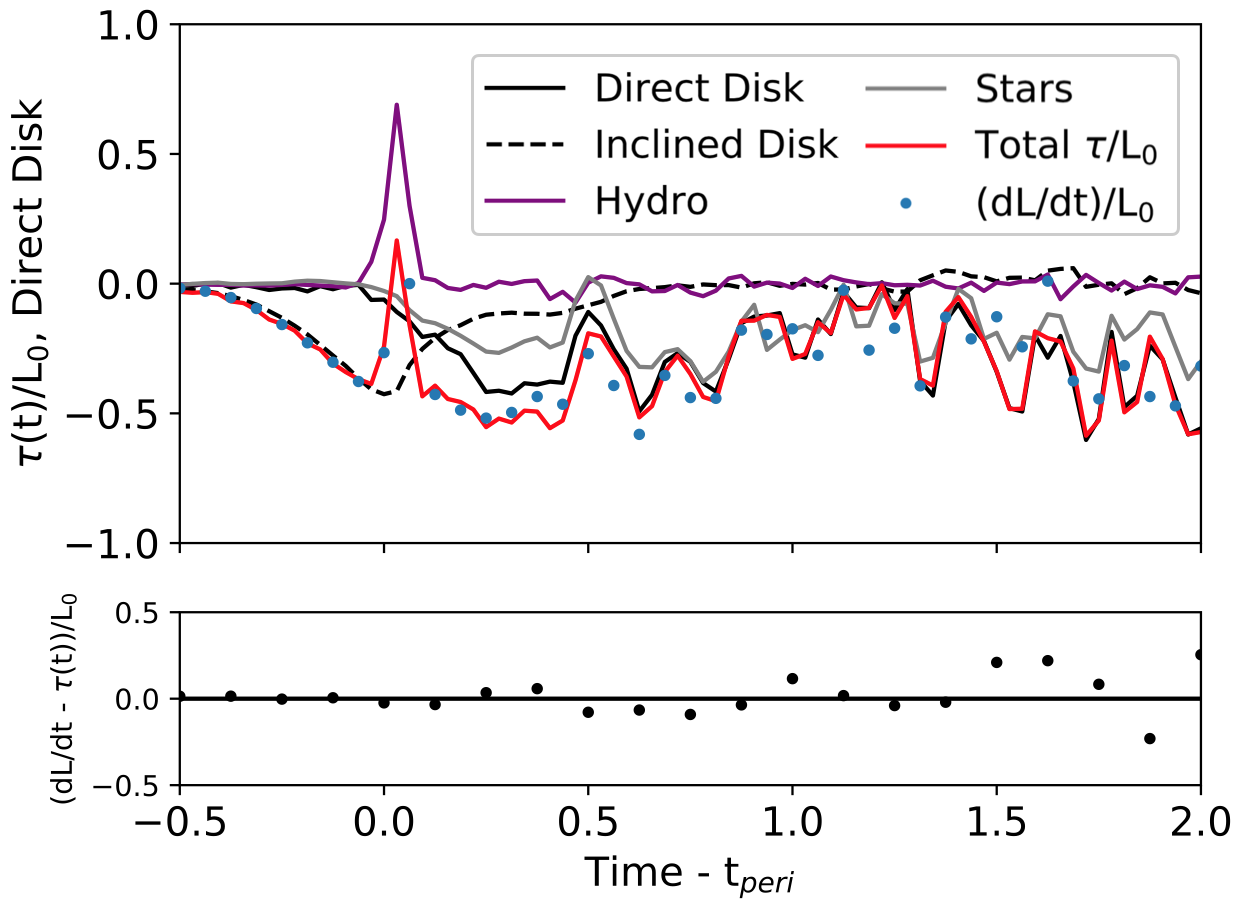}
	    \caption{Measured and estimated torques on the direct disk of our canonical encounter: D1. This shows the total torque in red, which is made up of components due to the direct galaxy (black), the inclined galaxy (black dashed), and the hydrodynamic force (purple). Residuals, or the difference between the measured total torque (i.e., including all gravitational and hydrodynamic forces) and the time derivative of the angular momentum, are displayed immediately below the corresponding torque plot.}
		\label{fig:torque_example}
	\end{center}
\end{figure}

Figure \ref{fig:torque_example} shows our key example: the torques on the direct disk of encounter D1. Here, and in subsequent plots, torques have been normalized by dividing values by the Lagrangian volume's initial angular momentum, L$_{0}$, implying that the vertical units on these plots are inverse time. The total torque, $\tau (t)$, calculated directly from the net gravitational and hydrodynamic forces acting on the Lagrangian volume, is shown in red. For comparison, we also plot the numerical derivative of the volume's angular momentum, d$L$/d$t$ (blue dots); residuals between $\tau (t)$ and d$L$/d$t$ are shown in the bottom panel. On the whole, the residuals are small and fluctuate around zero; this gives us confidence that our techniques for centring and calculating torque are correct (see Appendix \ref{sec:measuretorques} for details).

In addition, Figure \ref{fig:torque_example} also shows partial torqes due to the direct galaxy's gravity (solid black), the inclined galaxy's gravity (dashed black), the direct galaxy's stellar disk (grey), and the net hydrodynamic force (purple). This shows how various components conspire to drive gas inward. Near first encounter, angular momentum is lost to the companion by gravitational torques. In contrast, hydrodynamic forces at first pericentre briefly increase the angular momentum. By about 0.5 time units after pericenter, much of the material in the Lagrangian volume has fragmented into clumps which are losing angular momentum to their host stellar disk (and also the bulge and halo, as they have some nontrivial contributions to the overall torque on the gas) via gravitational interactions. Comparing back to Figure \ref{fig:inflow_d1}, we saw that there is almost no increase in nuclear material between first pericentre and $t-t_{peri}=1.5$, which is when the first clump reaches the centre.

\begin{figure}
	\begin{center}
	    \includegraphics[width=0.485\textwidth]{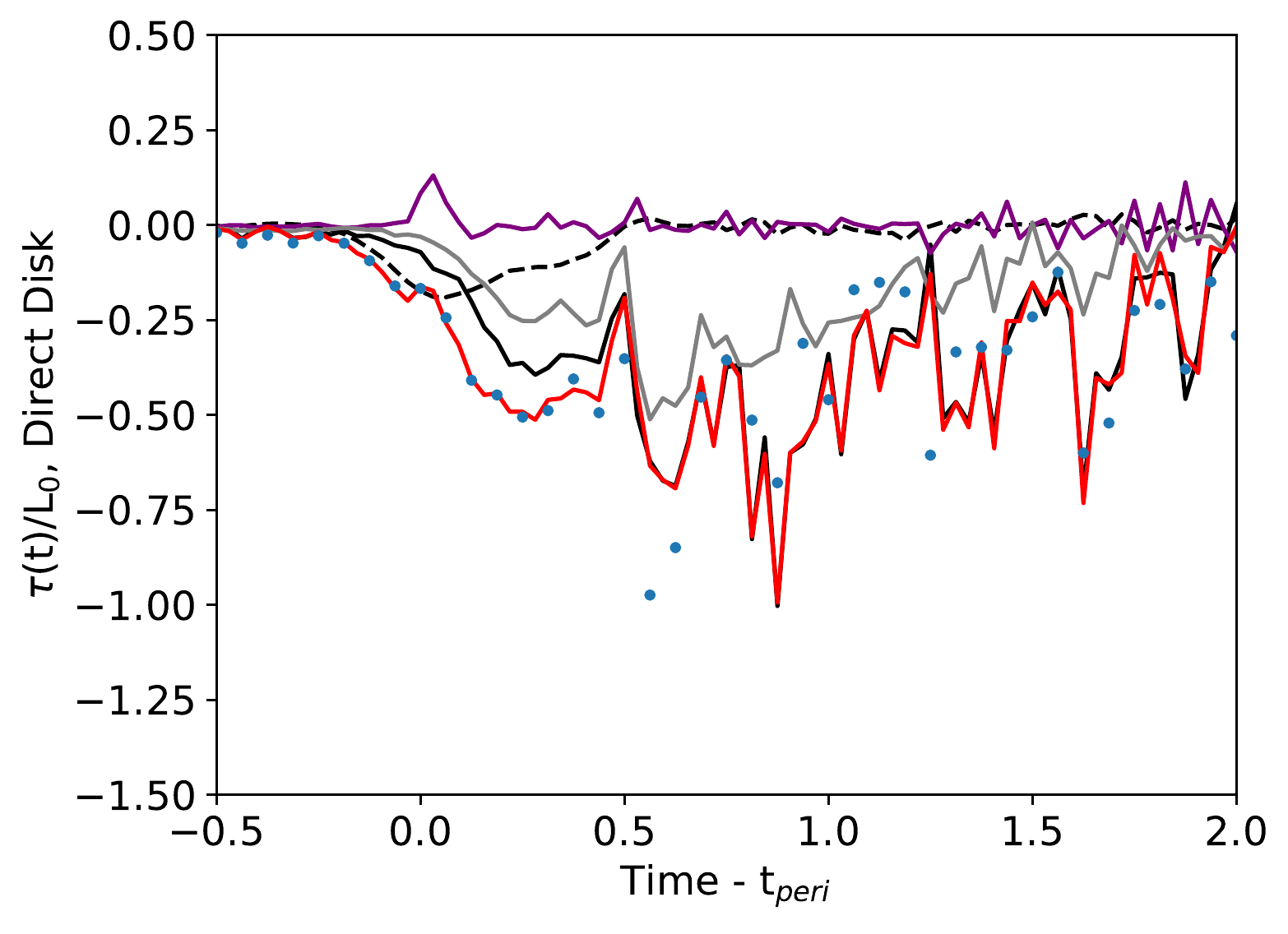}	
	    \caption{Here, we show the torques experienced by the Lagrangian volume in the direct galaxy of the A1 encounter.}
		\label{fig:torques_r013}
	\end{center}
\end{figure}

In \S \ref{sec:inflow} we noted that the direct disk in encounter A1  produces nearly twice as much inflow as the corresponding disk in encounter D1. Figure \ref{fig:torques_r013} shows the torques acting on the Lagrangian volume for the A1 encounter, and it is clear why the resulting inflow is so much larger. The initial hydrodynamic torque, which opposes inflow, is significantly weaker, and the magnitude of the direct disk's self-gravitational torque is larger in comparison. As in the example above, the total torque in the A1 direct disk appears to be dominated by stellar material early on. However, there are other sources of torque in operation, as the stellar disk accounts for only about half of the net torque measured at later times. 

\begin{figure}
	\begin{center}
	    \includegraphics[trim = 6.75cm 3.75cm 6.75cm 3.75cm, clip, width=0.5\textwidth]{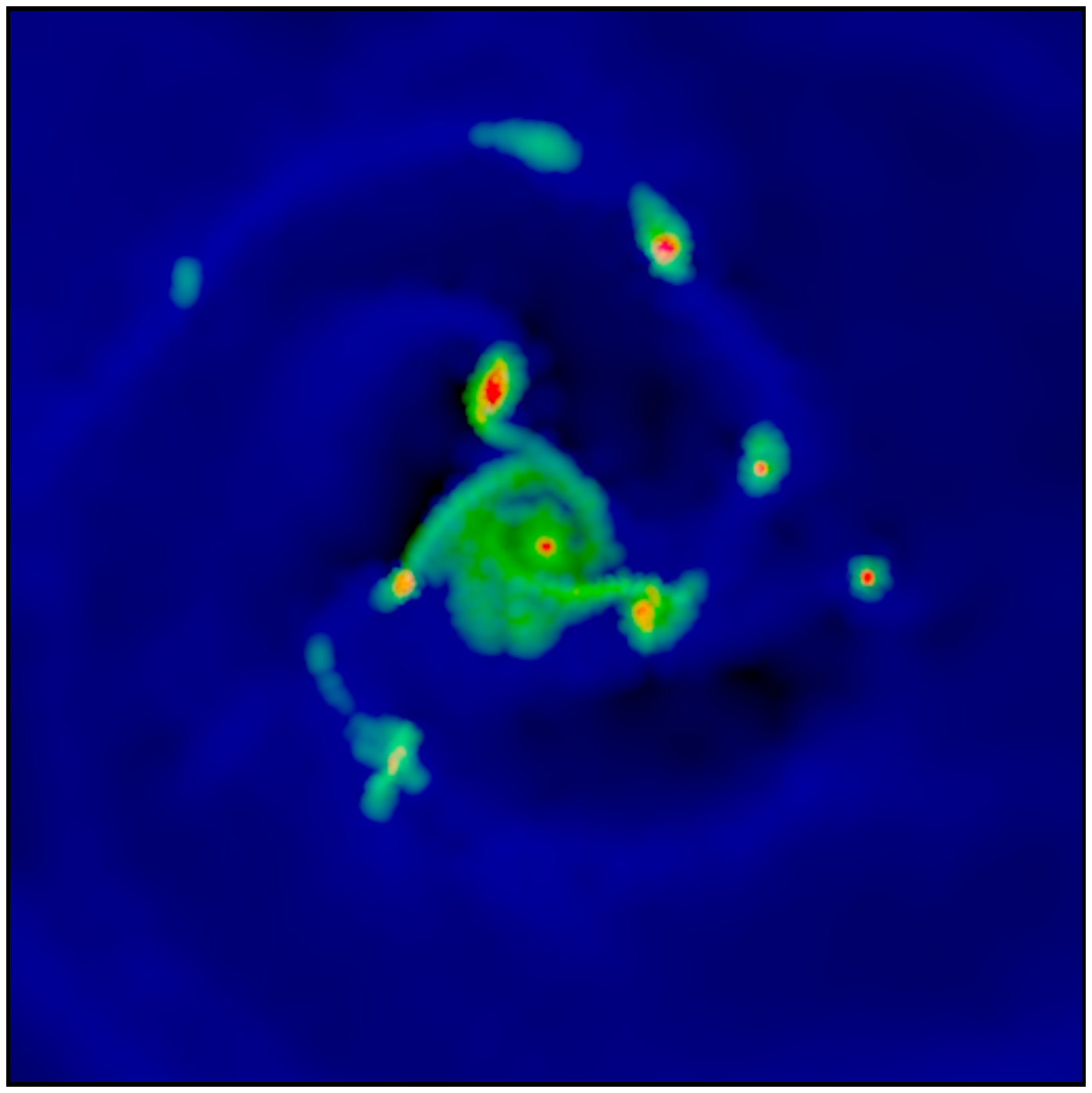}
	    \includegraphics[trim = 6.75cm 3.75cm 6.75cm 3.75cm, clip, width=0.5\textwidth]{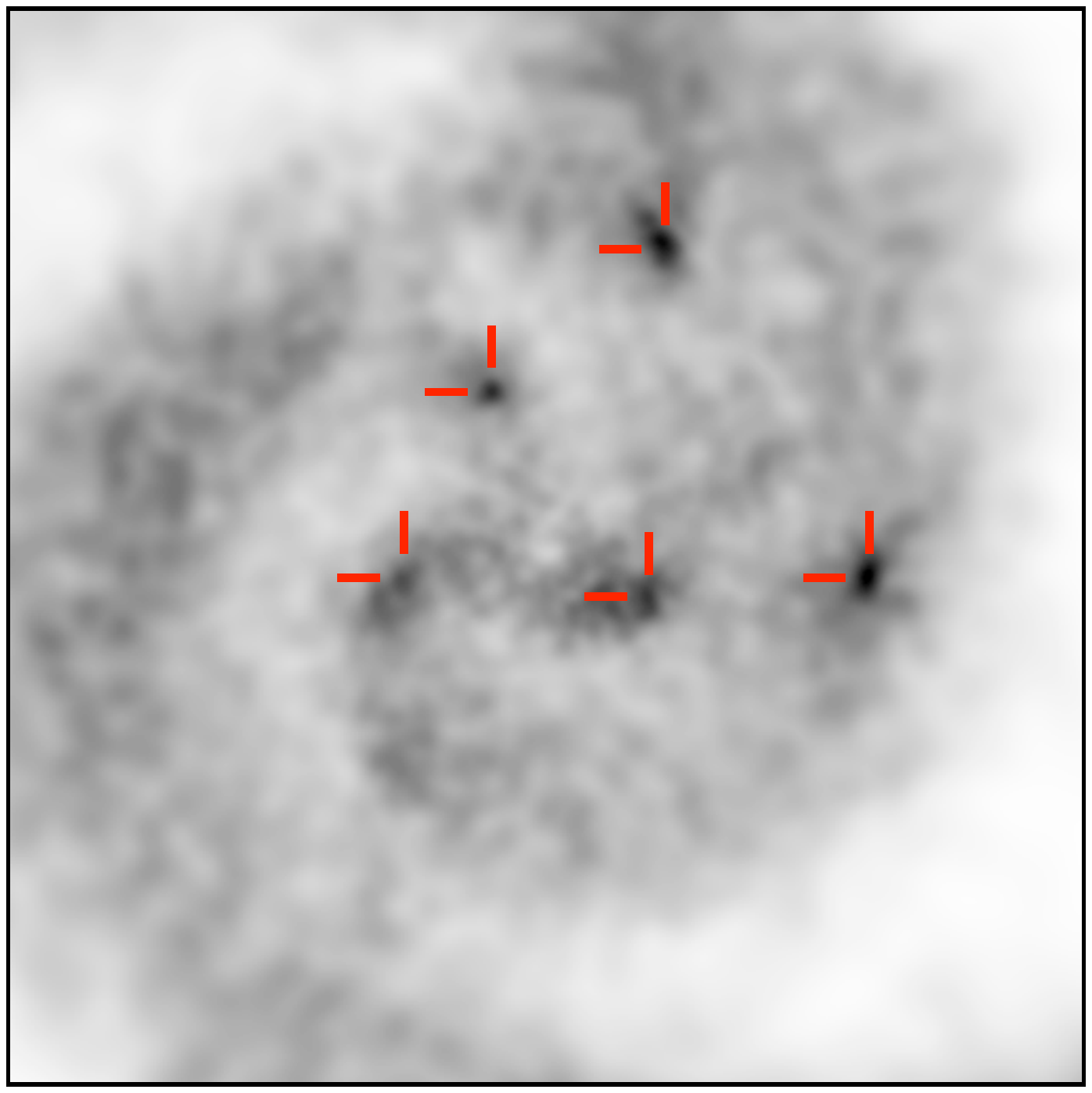}
	    \caption{A close-up of clump formation, corresponding to $t - t_{peri} =0.91$, with 0.25x0.25 length units on a side. \textit{Top:} Gas density map to show clump formation at this time. \textit{Bottom:} The stellar distribution, normalized as a function of radius to show density enhancements due to clump formation. Red crosshairs mark the location of the clumps in the gas density. The simulation represented here is fundamentally similar to A1, but with four times as many stellar particles to improve the stellar mass resolution. Note how clumps create localized overdensities in the stellar distribution. This is direct evidence of dynamical friction at work.}
		\label{fig:clumps}
	\end{center}
\end{figure}

Figure \ref{fig:clumps} provides a close-up of clump formation following a strong tidal interaction. Shocks form just after first pericentric passage, creating filaments of gaseous material which become Jeans-unstable. The resulting clumps orbit the centre of their host galaxy, and eventually spiral to the centre as a result of gravitational torques. Clumps have been seen before in simulations \citep[e.g.,][]{li2004, li2005}, but were observed in both isolated {(control)} and merging disks, indicating that those disks were less stable than the ones presented here. In this earlier work, the clumps were considered the progenitors of globular clusters. The clumps seen in our simulations are on average about 0.001 mass units, 0.0033 length units in radius and have lifetimes around 0.1 time units, {or in physical units, $m_{c} \approx$ 10$^{8}$M$_{\odot}$, $r_{c} \approx$ 100pc, and $t_{c}  \approx$ 25 Myr, roughly consistent with potential progenitors of globular clusters \citep[e.g.,][]{elmgreen2005}.}

Figures \ref{fig:torque_example} and \ref{fig:torques_r013} show explicitly that the hydrodynamic force is \textit{not} responsible for torquing material inward. Instead, the stellar distribution is responding to the gas clumps. In Figure \ref{fig:clumps}, it is clear to see that there is a wake of trailing stellar material behind the prominent gas clumps. Stellar overdensities {gravitationally drag the clumps, removing orbital energy and angular momentum, thereby forcing them to spiral toward the center. This is an example of dynamical friction, also discussed briefly in \cite{duc2004} as a way by which clumps may migrate inward.} Figures \ref{fig:torque_example} and \ref{fig:torques_r013} show that the torques measured from the stellar distribution are in fact the cause of the gas migrating toward the center. 

\subsubsection{Ram-Pressure Sweeping}
\label{sec:hydro}

\begin{figure}
	\begin{center}
	    \includegraphics[width=0.485\textwidth]{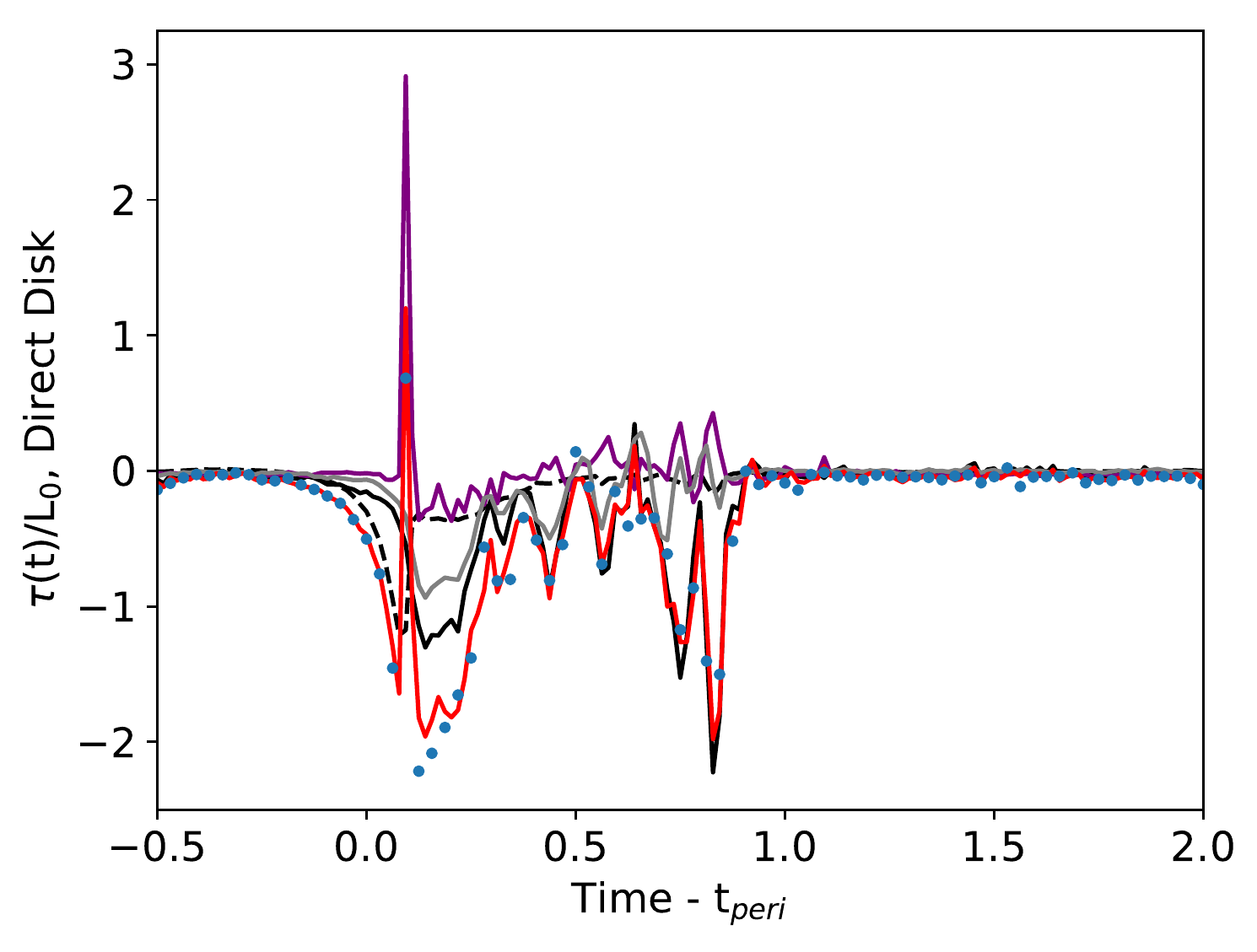}	
	    \caption{Similar to Figure \ref{fig:torques_r013}, but for encounter A4. The line and symbol styles match the legend of Figure \ref{fig:torque_example}. The evolutionary timescale is compressed since this relatively close encounter merges more rapidly than the other cases described here.}
		\label{fig:torques_r021}
	\end{center}
\end{figure}

A good foil to the A1 encounter is the closer A4 encounter, shown in Figure \ref{fig:torques_r021}. As Figure \ref{fig:inflow_d1} reports, the direct disk of this encounter has relatively little inflow, unlike D1 and A1. The much-closer passage results in a rather strong but very brief hydrodynamic torque, while at later times the gravity of the stellar component becomes the main source of torque. This suppresses inflow because the direct disk gains angular momentum as a result of hydrodynamic forces. Thus, it is undergoing a large ram-pressure effect.

\begin{figure*}
	\begin{center}
		\includegraphics[width=0.49\textwidth]{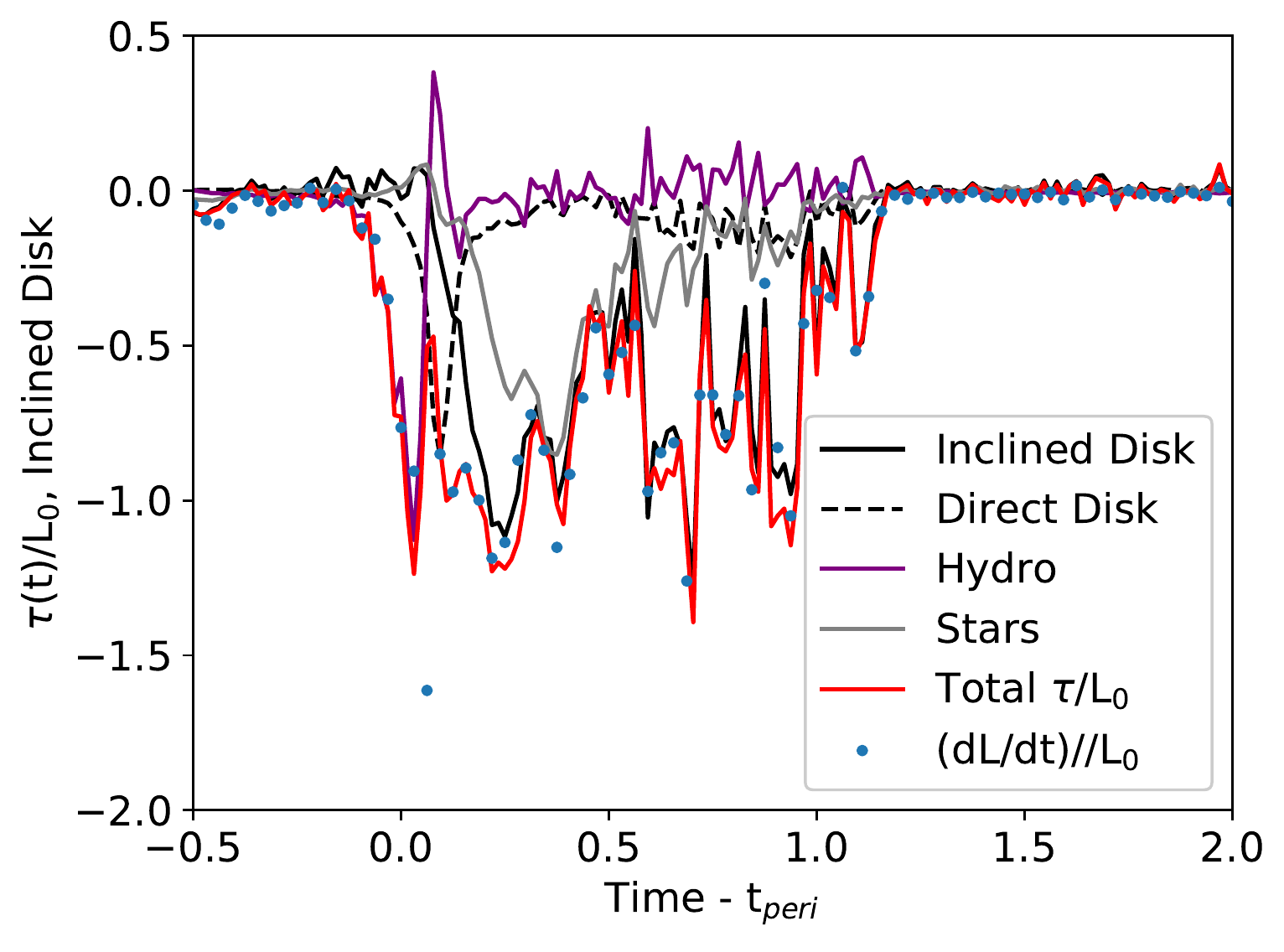}
	    \includegraphics[width=0.49\textwidth]{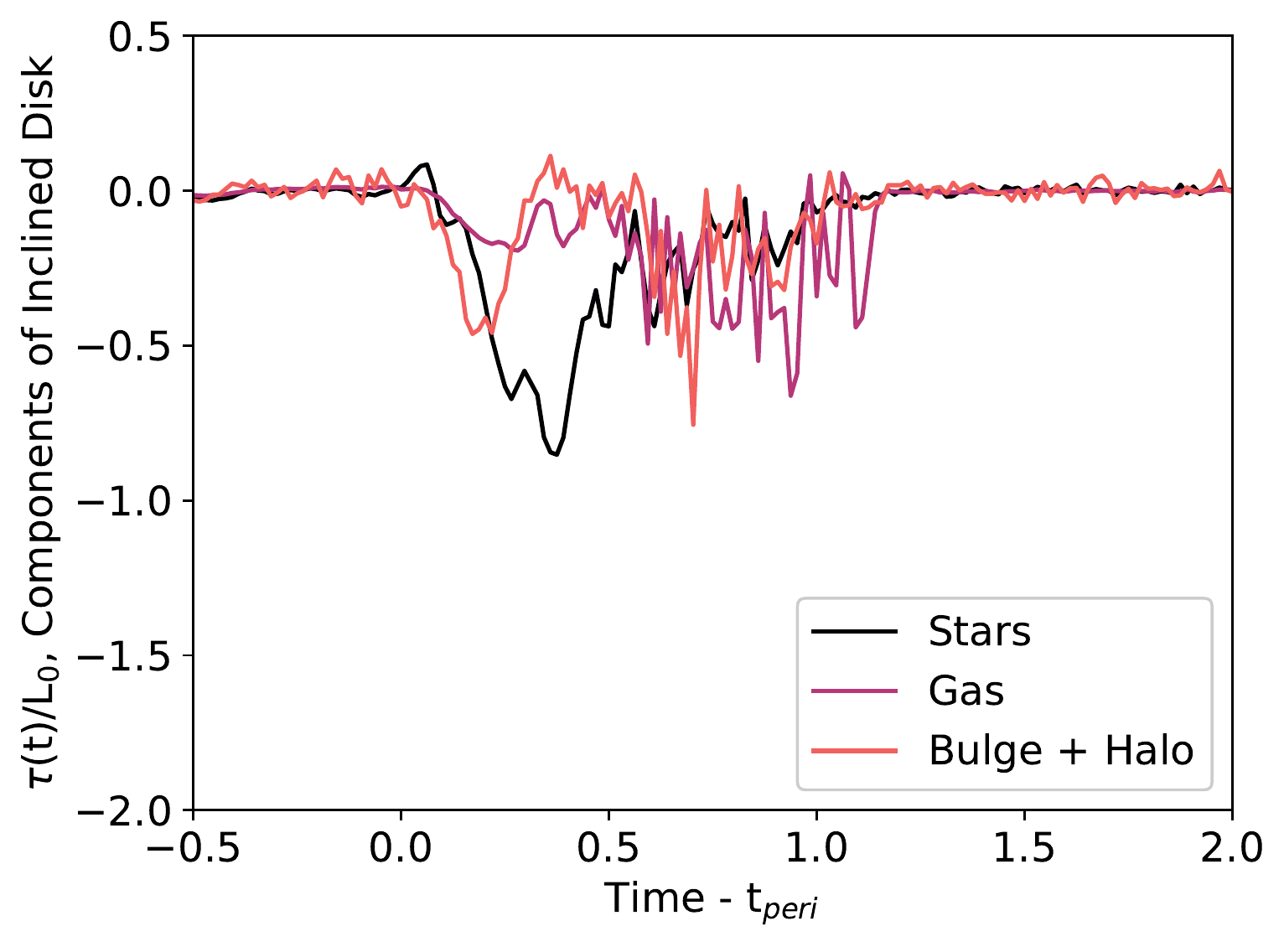}	    
	    \caption{This shows the torques on the inclined disk of encounter A4. In addition, the right panel shows the different components of the galaxy: stars, gas (gravitational), bulge and halo. With this, we can pick apart the main source of the torque.}
		\label{fig:d2torques_r021}
	\end{center}
\end{figure*}

Comparing back once more to Figure \ref{fig:inflow_d1}, we found that the inclined disk of encounter A4 experiences nearly three times as much inflow as the direct disk. As Figure \ref{fig:d2torques_r021} shows, the inclined A4 disk experiences a very strong hydrodynamic torque, which, in this particular geometry, \textit{subtracts} angular momentum and consequently transports gas inward. The right panel of Figure \ref{fig:d2torques_r021} analyzes the gravitational torques due to the individual components of the galaxy. We see that the stellar component dominates the torque after first pericentre, and is overtaken by the gaseous component at later times. This plot shows a large contribution due to the bulge and halo components of the galaxy's total torque, which drives more material into the nucleus over a long period of time compared to its direct counterpart. The strong initial hydrodynamic response of the inclined disk produces more clumps, which in turn spiral into the nucleus via dynamical friction{, and deliver a larger amount of material than in the direct A4 disk.} 

This is a good example of how the two disks in a given encounter may have significantly different inflow mechanisms. While the A4 direct disk's inflow was primarily driven by clumps (with a small contribution from ram-pressure sweeping), the A4 inclined disk inflow was almost completely a result of ram-pressure sweeping. However, neither the direct nor the inclined disk produced as much inflow as their counterpart in the widest passage. Studies such as \cite{dimatteo2007} have shown that the star formation efficiency is anti-correlated with pericentric passage, perhaps due to the negative impact of this ram-pressure effect on inflow production.

\begin{figure}
	\begin{center}
	    \includegraphics[width=0.485\textwidth]{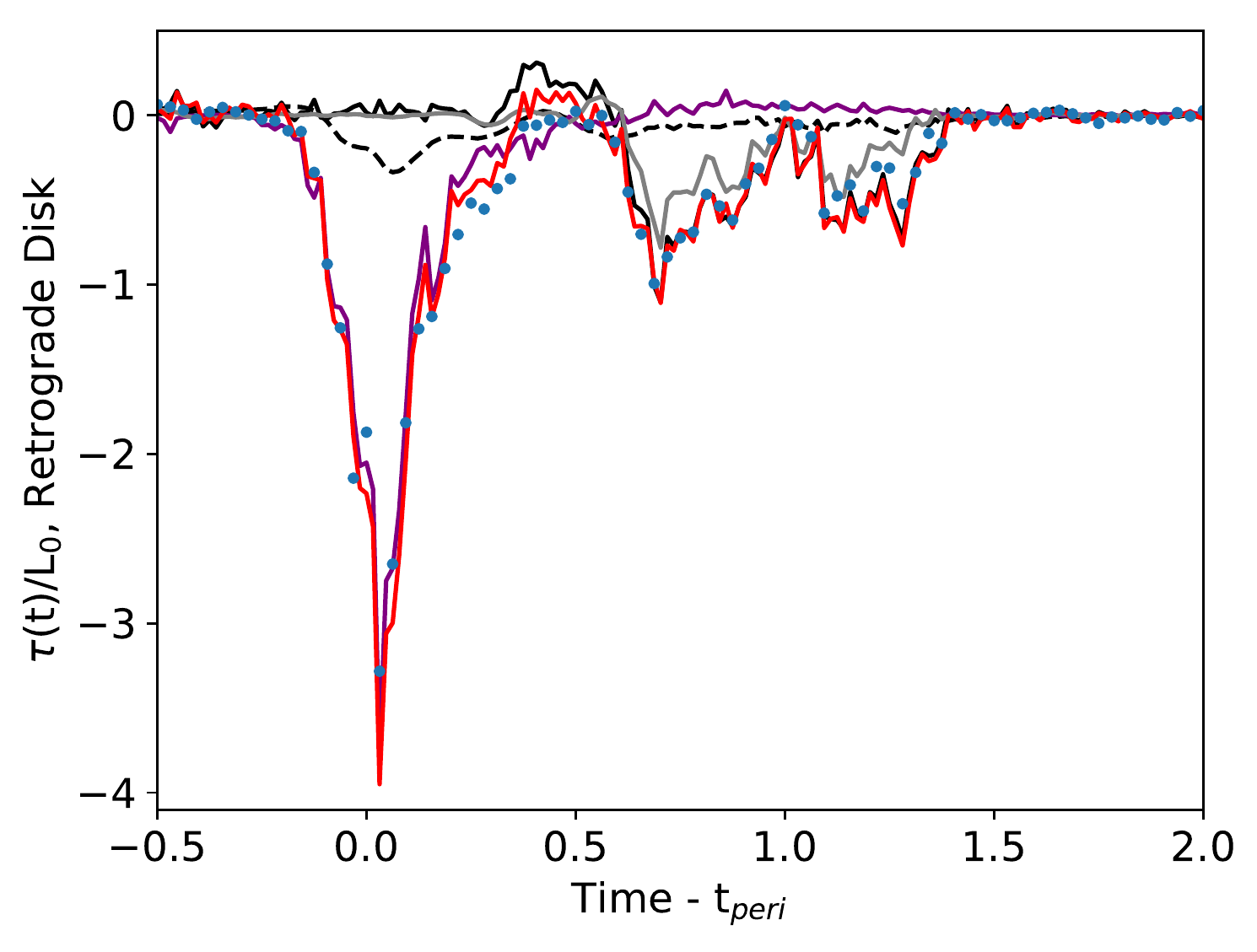}
		\vspace{12pt}
	    \caption{Similar to the left panels of Figures \ref{fig:torques_r013}, and \ref{fig:torques_r021}, but for the retrograde galaxy of encounter D1$_{r}$. The line and symbol styles match the legend of Figure \ref{fig:torque_example}. Note that the torque due to the hydrodynamic interaction dominates here, whereas in previous examples other components of the galaxy were the main drivers of inflow.}
		\label{fig:torques_r024}
	\end{center}
\end{figure}

{Ram-pressure sweeping is also at play in our} two retrograde encounters, A1$_{r}$ and D1$_{r}$. In the D1$_{r}$ case, the hydrodynamic torque resulting from the deep retrograde encounter of the two extended gas disks is responsible for over half of the angular momentum loss, as can be seen in Figure \ref{fig:torques_r024}. The geometry of the gas disks enables a strong hydrodynamic interation, which in turn initiates the encounter-induced inflow.  This phenomenon has been noted before, as an alternative route to nuclear fueling \citep[e.g.,][]{capelo2017, barnes2002}. In the case of D1$_{r}$, dynamical friction becomes important at late times; the gravitational torques are acting on a rather large gas clump which forms in the immediate aftermath of the encounter. 

It is worth noting that hydrodynamic torques often play a supporting role in transporting gas inward in inclined (as in A4) and/or retrograde disks. In inclined prograde encounters (e.g, Figure \ref{fig:d2torques_r021}), we find hydrodynamic forces initially acting to move gas inward, while at later times, gravitational torques predominate. While these initial torques are not strong enough to drive gas all the way to the nucleus, they transport gas to smaller radii where it can more effectively couple to the stellar disk, creating conditions which then enable gravitational torques to drive gas into the nucleus. Hydrodynamic torques can also work in the opposite sense in direct disks, preventing material from flowing to the center via a transfer of angular momentum from one disk to the other.

\subsubsection{Mode-Driven Inflow}
The right panel of Figure \ref{fig:bars_clumps} shows a familiar scenario \citep[e.g.,][]{barnes1991, mihos1996}: {modes} form as a result of the interaction, and rapidly drive inflow to the centre of the host galaxy. {Modes are produced by any non-axially symmetric structure. As an example, barred systems produce} organized large-scale flows which misalign the stellar and gaseous bars and consequently generate a torque on the gas. The stellar bar acts to slow down the gas bar, which in turn draws gas inwards.

\begin{figure}
	\begin{center}
	    \includegraphics[width=0.485\textwidth]{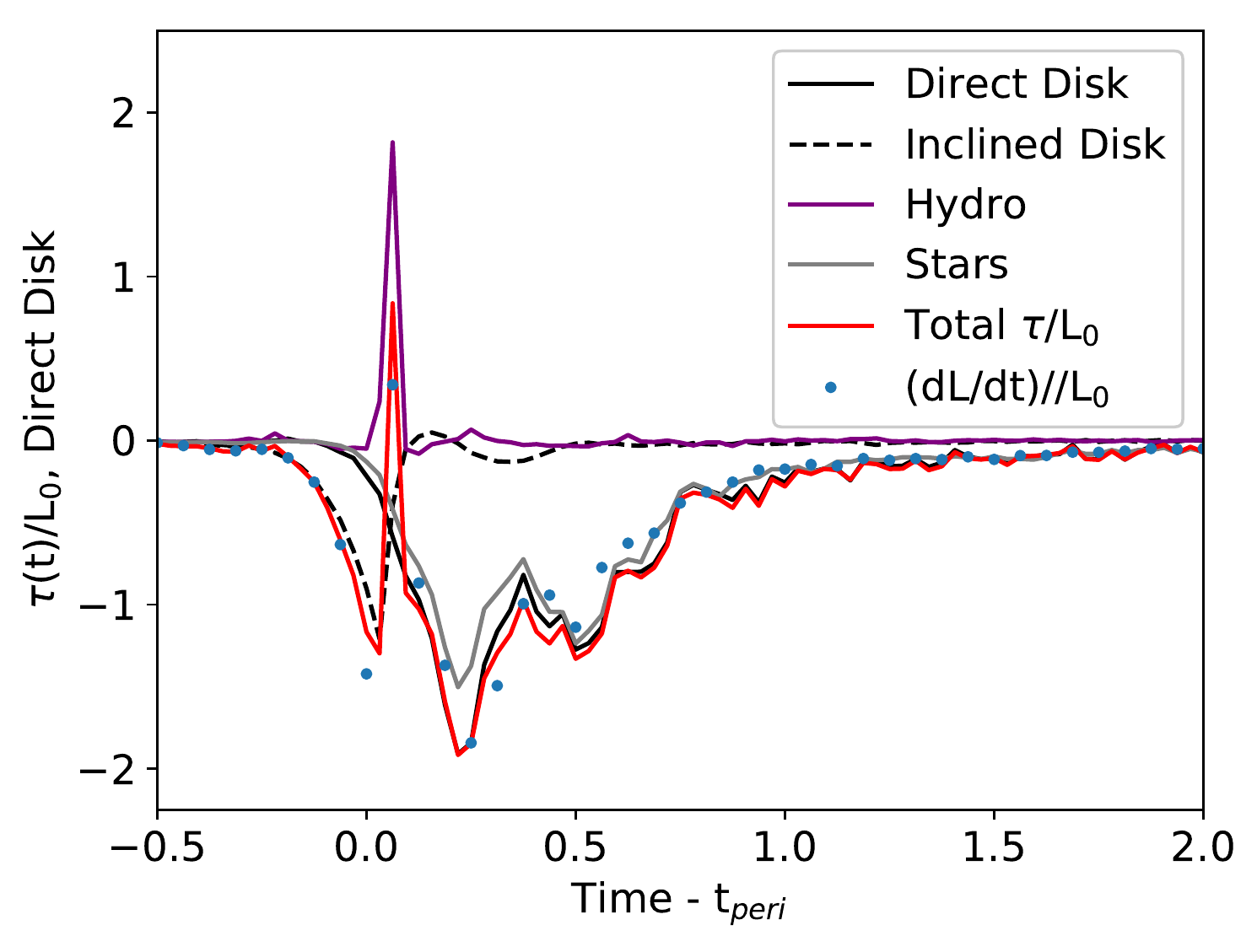}
	   	    \caption{Similar to Figures \ref{fig:torques_r013} and \ref{fig:torques_r021} direct galaxy of encounter E3'. Bars form as a result of this encounter.}
		\label{fig:torques_r125}
	\end{center}
\end{figure}

In Figure \ref{fig:inflow_d2}, we saw that these {mode-driven} inflows are rapid and continuous, contrasting with the arrival of clumps seen in Figures \ref{fig:inflow_d1} - \ref{fig:inflow_retro}. The legacy models {are thus mode-dominated, and induce continuous flow,} instead of fragmenting into gas clumps. Figure \ref{fig:torques_r125} shows the torques on the direct disk of the E3' encounter. The stellar component dominates the total torque after first pericentre, but prior to that time, the inclined disk significantly torques the direct disk. Figure \ref{fig:inflow_d2} also showed that there is a significant difference in inflow between the direct and inclined galaxies, similar to the reaction of the A4 encounter.  This is because the inclined disk does not experience the hydrodyanmic impulse. 

\section{Discussion}
A great deal of the development of our understanding of gas dynamics in the last two decades has been based on isothermal models \citep[e.g.,][]{hernquist1989, barnes1991}. This work has picked up where they left off. It is important to realize, however, that these simulations are limited in multiple ways: resolution effects; the treatment of the ISM equation of state; and the absence of additional subgrid physics including star formation and feedback. In this section, we will address these limitations {and provide a deeper context for our simulations.}

{
\subsection{Previous Descriptions of Clumps}
It is important to note that this is not the first time clumps have been observed in simulations. In work such as \cite{dimatteo2007}, the authors see clumps form, and mention that these structures are the cites of off-nuclear star formation. However, the authors do not attempt to quantify these clumps in any way. Clump formation has been studied within the context merger simulations \citep[e.g.,][]{dekel2009, tey2010, bour2011, renaud2014}, but never to the level of detail as discussed here. For example, \cite{dekel2009} found that their clumps contain a significant fraction of the gas mass, and convert to stars on roughly the same time scale that they migrate in on, thus building up the bulge. This does imply that the clumps, at least in their simulations, survive the inclusion of star formation and stellar feedback.  \cite{renaud2014} attempted to connect the structure of the interstellar medium to inter-galactic dynamics, and concluded that a clumpy interstellar medium is a byproduct of turbulence. \cite{bour2011} went one step further and included clumps into their initial gas distributions, in an attempt to accurately represent encounters of galaxies at high redshift, which are often observed to be clumpy \citep[e.g.,][]{elmgreen2005, cev2010}.}

\subsection{Resolution Considerations}
\label{sec:res}
Smoothing a gravitational potential serves to weaken that potential. Similarly, altering the number of gas particles included within a single SPH kernel changes the pressure that each particle exerts. Early work, such as \cite{bate1997}, stressed the importance of matching these two resolutions when dealing with Jeans fragmentation.

The Jeans length can be written as
\begin{equation}
\lambda_{J} = c_{s} \sqrt{\frac{\pi}{G}} \rho^{-1/2}
\end{equation}
where $c_{s}$ is the sound speed, given by $c_{s} = \sqrt{(\gamma - 1)u_{int}}$, and $\rho$ is the SPH density. Similarly, the Jeans mass can be given by the expression
\begin{equation}
M_{J} = \frac{4\pi}{3}\rho\left(\frac{\lambda_{J}}{2}\right)^{3}  = \frac{\pi^{5/2}}{6} \frac{c_{s}^{3}}{G^{3/2}\rho^{1/2}}
\end{equation}
By equating $\lambda_{J}$ to the gravitational smoothing length, $\epsilon$, and $M_{J}$ to the mass of an SPH smoothing volume, $40M_{SPH}$, we can derive critical densities, $\rho_{crit}$, at which Jeans fragmentation can be resolved. If the $\rho_{crit}$ derived from $\lambda_{J}=\epsilon$ is smaller than that derived from $M_{J}=40M_{SPH}$, then fragmentation will dominate the disk. On the other hand, if the $\rho_{crit}$ derived from $M_{J}$ is smaller, then fragmentation will be suppressed, resulting in a more (perhaps artificially) smoothed gas distribution. Ultimately, the relative magnitude of the $\rho_{crit}$'s derived from $\lambda_{J}$ and  $M_{J}$ depends on the science goal.

Using $\epsilon = 0.0025$, $40M_{SPH} = 3.815 \times 10^{-5}$, $\gamma = 5/3$ (as for a monatomic gas) and $u_{int} = 0.014$, we can derive the critical densities for the majority of our simulations. We find that
\begin{equation}
\rho_{crit, grav} \approx 4690 \> \> \> \> \> \> \> \> \> \> \> \> \mathrm{and} \> \> \> \> \> \> \> \> \> \> \> \> \rho_{crit, SPH} \approx 4750
\end{equation}
The insight to be gained here is that there is a minimum mass and size scale that we can resolve given the fluid element mass and smoothing length chosen. The characteristic scale of clumps derived from these two quantities is likely a lower-limit to the size of clumps produced in our simulations. Thus, clump formation {and its eventual migration is likely} a generic mechanism, but the characteristic scales are sensitive to the resolution choices we make in the simulations. 

{\cite{tey2010} posited that large-scale inflows of gas were in fact an artifact of a poorly resolved interstellar medium. }More recent work \citep[i.e.,][]{maji2017} has investigated how star cluster properties differ in SPH codes and meshless codes. They also discuss the limitations of SPH codes, and the importance of considering the resolution when interpreting results.

Inadequate mass resolution in the stellar disk may potentially influence the formation of clumps \citep[e.g.,][]{wetz2007}. We have run one version of encounter A1 with four times the number of gas particles, and another version with four times the number of stellar particles, to test the effect of improving both the gas and stellar mass resolutions. Not only do clumps still form in these simulations, but the improved mass resolution of the stellar distribution makes it much easier to identify how the clumps migrate into the center of the galaxy. Density fluctuations in the stellar distribution trail behind clumps in the gas, implying that the clumps we see are indeed being driven into the nucleus via dynamical friction. This does not necessarily contradict the \cite{wetz2007} work, as they were studying the creation of clumps in tidal tails, which is a much more rarefied environment than the disk of an interacting galaxy.

\subsection{Missing Subgrid Physics}
There has been a significant amount of work recently \citep[e.g.,][]{mihos1996, li2004, chien2010, hopkins2013, hay2014, li2014a, li2014b, beh2016, kim2017, mand2017, oklop2017} on the effects of star formation and feedback on gas disks, both in isolation and in mergers. For example, \cite{hopkins2013} found that much of the gas in their resolved giant molecular clouds (GMC's) transitioned to stars, leaving a small amount of gas still gravitationally bound to the GMC to spiral in toward the center. It may very well be that the clumps we see in this work would dissipate as a result of star formation. This could result in off-nuclear star formation, for which there is a wealth of observational evidence \citep[e.g.,][]{hag2007, evans2008, cor2011}. If these clumps do wind up staying bound as star clusters, then they may spiral inward via the same process discussed here to get bulge growth, without having to form stars at the nucleus. Any remaining gas within these clusters could also fuel AGN.

High redshift galaxies are known to be very clumpy, perhaps as a result of violent disk instability \citep[][]{dekel2009}. \cite{mand2017} looked at high redshift disks in cosmological simulations, with and without feedback from radiation pressure, which serves to reduce the amount of gas available to star formation. They found that this kind of feedback reduces the mass of the clumps by nearly an order of magnitude, and thus reduces their lifetimes significantly. \cite{kim2017} attempted to locate candidates for globular clusters within the FIRE simulation. They suggest that mergers at high redshift might be the perfect environment for these objects to form. 

\section{Conclusions}
\begin{figure*}
	\begin{center}
	    \includegraphics[trim = 1cm 0cm 2cm 0cm, clip, width=0.7\textwidth]{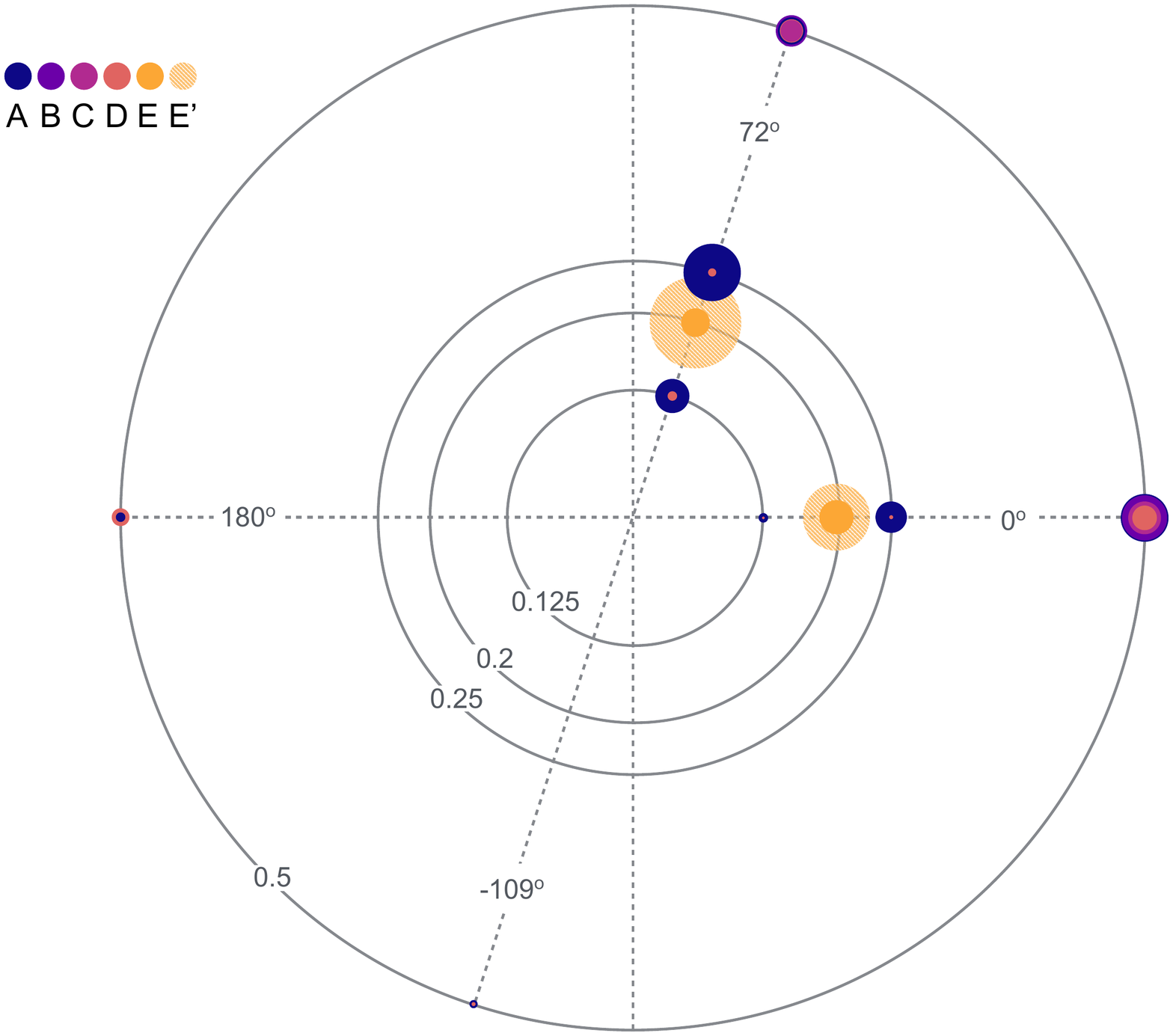}
	    \includegraphics[trim = 2.5cm 2cm 2.5cm 1.5cm, clip, width=0.7\textwidth]{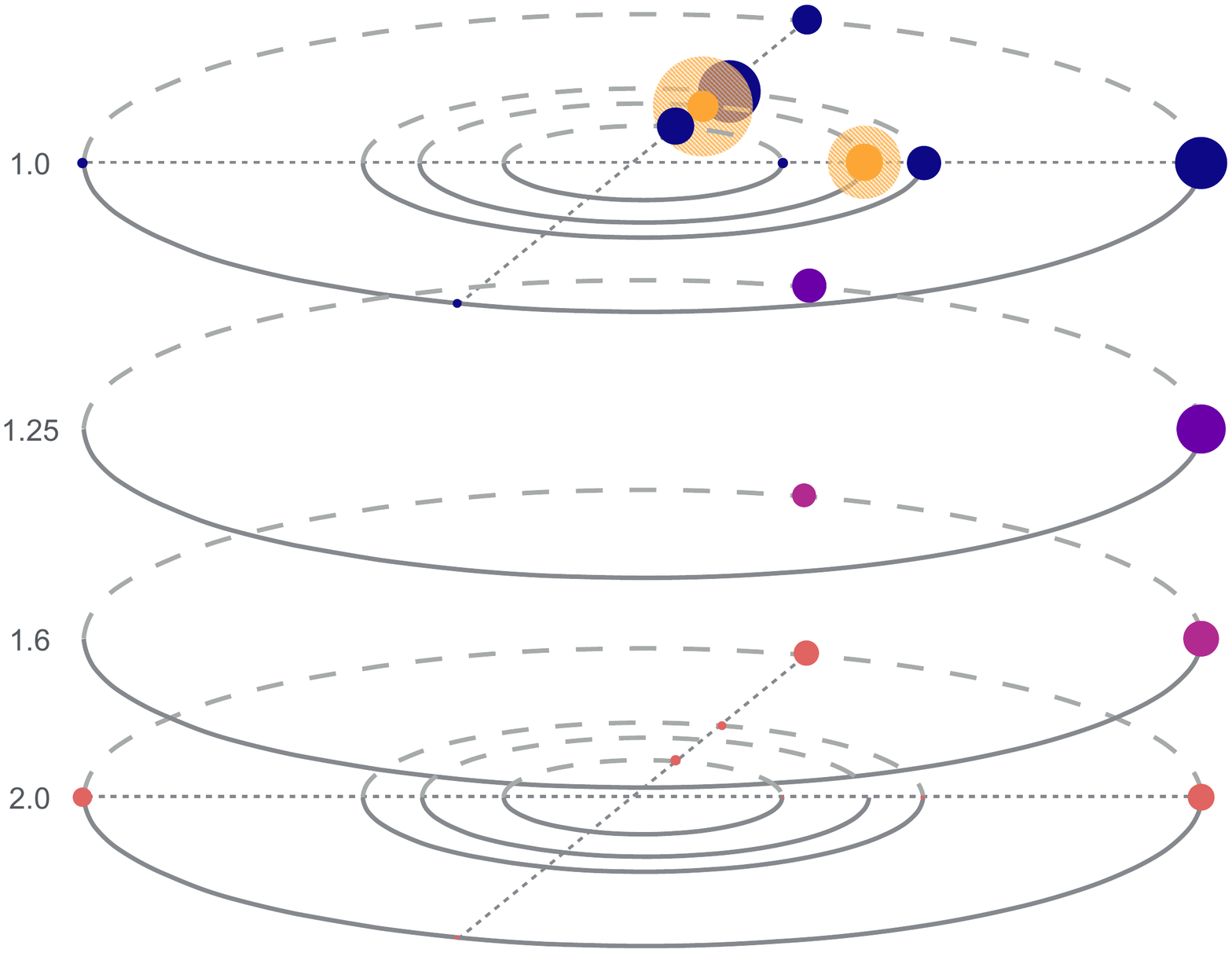}	 
	    \caption{Here we graphically summarize our results. Concentric circles show the range of pericentric separations; the inclination angles are indicated by their position on the unit circle. The diameter of each point corresponds to the final nuclear mass of each run. The bottom figure shows a side-view of the top; each layer corresponds to a different $\alpha_{\star}/\alpha_{g}$.}
		\label{fig:sumplot}
	\end{center}
\end{figure*}

In this paper, we have disclosed multiple paths by which gas can arrive at the centre of galaxies as a result of major mergers. We highlight the different mechanisms which drive inflow by exploring the parameter space of encounters. Figure \ref{fig:sumplot} summaries the results of our many simulations graphically. Concentric circles show the range of pericentric separations; the inclination angles are indicated by their position on the unit circle; the diameter of each point corresponds to the final nuclear mass of each run; the color of each point corresponds to the mass model indicated. The bottom panle shows a side-view of the top; each layer corresponds to a different $\alpha_{\star}/\alpha_{g}$. From this, we can clearly see that there is an intricate relationship between inclination angle, pericentric separation, and resulting nuclear inflow. Further, this relationship is dependent upon mass model, gas fraction and gravitational smoothing. Our results pose important implications for how interactions might affect nuclear activity. Our conclusions are as follows:
\begin{enumerate}
\item From the outset, we expected to see bars form and drive material into the nucleus. Our original goal was to find out what happens to the inflow when the stellar bar is much smaller than the gas disk. However, we found that the {modal} mechanism is not the only way to get nuclear inflows, and may not be the most important way. Shock fronts form just after first pericentric passage, producing filaments of gaseous material which become Jeans-unstable, eventually forming massive, dense clumps. In our simulations, these appear to be the main drivers of material into the central regions of the galaxy and do so via dynamical friction. 
\item Simulations such as those in \cite{barnes1991} posited that bars were the main drivers of inflow; however we have found that the bar formation in that case was in fact a result of a high smoothing parameter and low gas fraction. When {non-axially symmetric structures, such as bars} do form, it is highly dependent upon gas fraction (that is, lower gas fraction typically leads to the formation of a stronger bar). 
\item The efficiency of inflow is intricately dependent upon the encounter geometry. That is, the size of the gas disc, inclination of the encounter, and pericentric separation all conspire to affect the process of inflow. For prograde encounters, inflow is most efficient in small disks that come in on wide orbits. Retrograde encounters seem to have more efficient inflow in a large gas disc. 
\item Clump driven inflow and {mode-}driven inflow have different timescales for nuclear delivery. Bars promptly deliver a large amount of gas, while clumps intermittently deliver material over a longer period of time.
\item Encounter geometry is a very important factor for inflow. Depending on the circumstances, inclined disks may actually experience a larger inflow than direct disks, contrary to what was previously understood \citep[e.g.,][]{barnes1996, mihos1996}.
\item Inflow is strongest when the gas and stellar disks are the same size. More extended gas disks retain too much of their original angular momentum and are not well coupled to the stellar material, thus  providing less inflow. It is therefore plausible that previous simulations using stellar and gas disks of comparable size have over-estimated the amount of inflow, and perhaps, the star formation rate.
\end{enumerate}

\section*{Acknowledgements}
This material is based upon work supported by the National Science Foundation Graduate Research Fellowship under Grant No. DGE-1329626. The authors would also like to acknowledge the aid of both anonymous referees, Lars Hernquist, Brent Tully, and Larissa Nofi in preparing this document for publication. 

\bibliographystyle{mnras}
\bibliography{arxiv_bib}{}


\begin{appendix}
\section{SPH Code}
\label{sec:sphcode}
The SPH code used for these simulations incorporates algorithms from TREESPH \citep[][]{hernquist1989} and Gasoline
\citep[][]{wad2004}; gravitational forces are calculated using a Tree algorithm \citep[][]{barnes1986}. Previous studies using this code include \cite{barnes2002, barnes2004, chien2010}.

As in other SPH codes \citep[e.g.,][]{mona1992}, gaseous material is represented using discrete particles, and kernel
interpolation is used to obtain hydrodynamic variables as functions of position.  Gas particles are subject to ``hydrodynamic'' forces due to pressure gradients as well as shocks; the net acceleration of gas particle $i$ due to such forces is
\begin{equation}
  \left( \frac{\nabla P}{\rho} \right)_i^\mathrm{SPH} =
    \sum_{j \ne i} m_j \,
      \left(\frac{P_i}{\rho_i^2} + \frac{P_j}{\rho_j^2} + \Pi_{ij}\right)
      \, \hat{\mathbf{r}}_{ij} \, \left. \frac{d\overline{W}}{dr}\right|_{r=r_{ij}} \, ,
\end{equation}
where 
\begin{equation}
\frac{d\overline{W}}{dr} = \frac{1}{2}\left[ \frac{dW(r,h_{i})}{dr} + \frac{dW(r,h_{j})}{dr}\right]
\end{equation}
and our notation follows \cite{mona1992} throughout.  Likewise, the internal energy $u_i$ of gas particle $i$ is subject to $PdV$ work and shock dissipation:
\begin{equation}
  \dot{u}_i^\mathrm{SPH} =
    \sum_{j \ne i} m_j \,
      \left(\frac{P_i}{\rho_i^2} + \frac{1}{2} \Pi_{ij}\right)
      \, \hat{\mathbf{r}}_{ij} \cdot \mathbf{v}_{ij} \left.\frac{d\overline{W}}{dr}\right|_{r=r_{ij}} \,
      \, .
\end{equation}

If the internal energy $u_i$ is allowed to vary according to Eq.~(2), the net energy of a self-gravitating SPH system is
\begin{equation}
  E_\mathrm{net} = U_\mathrm{grav} + T_\mathrm{kin} + E_\mathrm{int}
  \, .
\end{equation}
where $U_\mathrm{grav}$ and $T_\mathrm{kin}$ are the gravitational and kinetic energy, respectively, of the particle system, and
\begin{equation}
 E_\mathrm{int} = \sum_{i} m_i \, u_i
\end{equation}
is the internal energy of the gas particles.  In the absence of numerical errors, $E_\mathrm{net}$ is conserved.  As a slight abuse of
terminology, we refer to such an SPH system as ``adiabatic'' even though Eqs.~(B1) and~(B2) include non-adiabatic processes (shocks).

When implementing an \textit{isothermal} SPH system, it's tempting to ignore Eq.~(B2) altogether and simply set $u_i = \mathrm{constant}$. In this case, however, the quantity $E_\mathrm{net}$ defined in Eq.~(B3) is no longer conserved, and variations in $E_\mathrm{net}$ cannot be used to diagnose numerical errors.  Instead, we imagine that each gas particle is coupled to an external reservoir with constant temperature $T$ and infinite heat capacity.  Any $PdV$ work or shock heating is then transferred to the external reservoir, and the internal energy of gas particle $i$ obeys
\begin{equation}
  \frac{d u_i}{d t} =
    \dot{u}_i^\mathrm{SPH} + \dot{u}_i^\mathrm{EXT}
  \, ,
\end{equation}
where $\dot{u}_i^\mathrm{EXT} = - \dot{u}_i^\mathrm{SPH}$ represents energy exchanged with the external reservoir by particle $i$; thus $d u_i/d t = 0$ and the gas is isothermal.  The net energy is then
\begin{equation}
  E_\mathrm{net} =
    U_\mathrm{grav} + T_\mathrm{kin} + E_\mathrm{int} + E_\mathrm{ext} 
  \, ,
\end{equation}
where $E_\mathrm{ext}$, the energy in the external reservoir, obeys
\begin{equation}
  \frac{d E_\mathrm{ext}}{d t} = \sum_{i} m_i \, \dot{u}_i^\mathrm{EXT} \, .
\end{equation}
The net energy $E_\mathrm{net}$ is conserved and variations can be used to detect numerical errors.

\section{Measuring Torques}
\label{sec:measuretorques}
As with the inflow calculations, we took the first 8192 bulge or stellar disk particles (where the particles are sorted by binding energy) to define the positional centroid. However, because torque measurements depend on both position and acceleration, we also determined the acceleration centre using the first 8192 halo particles. If the bulge particles are used to determine the acceleration centre, there is a substantial amount of jitter in the centroid's motion. Similarly, the halo particles do not accurately track the positional centroid of the potential well because the halo particles are diffuse. In all encounters, this method tracks the nuclear material well, and minimizes the motion of the potential's position in phase space.

Torque is the result of a force, $\bar{F}$, applied to a lever arm, $\bar{r}$,
\begin{equation}
\bar{\tau} = \bar{r} \times \bar{F}
\end{equation}
This then produces a change in the angular momentum
\begin{equation}
\bar{\tau} = \frac{\mathrm{d}\bar{L}}{\mathrm{d}{t}}
\end{equation}

Using the centroiding described above, we calculated the angular momentum at each time step, and took the time derivative of that function to arrive at a derived torque. If we had calculated explicitly the acceleration due to each component (i.e., bulge, gas and stellar disks, halo, etc.) at every time step, then the calculated torque (i.e., the result of the $\bar{r} \times \bar{F}$ calculation) and the derived torque would be the same. However, this was not the case. Hence, there is some residual scatter between these two measurements, as seen in Figure \ref{fig:torque_example}. Instead, to calculate the force due to each component on the Lagrangian volume, we set up acceleration calculations using the position and velocity information of all particles at each time step, weighted such that we single out individual components. To account for precession, we align the torque and angular momentum vectors. 

\end{appendix}

\bsp	
\label{lastpage}
\end{document}